\documentclass[amssymb,aps,prb,floats,twocolumn,showpacs,superscriptaddress]{revtex4-1}
\usepackage{graphicx,color}% Include figure files
\usepackage{dcolumn}% Align table columns on decimal point
\usepackage{bm}% bold math
\usepackage{natbib}% bold math
\usepackage{siunitx}
\usepackage{textcomp}
\usepackage{tikz}
\usepackage{xcolor}

\usepackage{hyperref}

\definecolor{mathematicablue}{RGB}{94, 129, 181}
\definecolor{mathematicabrown}{RGB}{197, 110, 26}
\definecolor{mathematicagreen}{RGB}{85, 85, 85}

\begin{document}

\title{
Laser-induced ultrafast demagnetization  
and perpendicular magnetic anisotropy reduction 
in a Co$_{88}$Tb$_{12}$ thin film with stripe domains}

\author{M. Hennes}
\thanks{These two authors contributed equally}
\email{marcel.hennes@sorbonne-universite.fr}
\affiliation{Sorbonne Universit\'e, CNRS, Laboratoire de Chimie Physique -- Mati\`ere et Rayonnement (LCPMR), 75005 Paris, France }

\author{A. Merhe} 
\thanks{These two authors contributed equally}
\affiliation{Sorbonne Universit\'e, CNRS, Laboratoire de Chimie Physique -- Mati\`ere et Rayonnement (LCPMR), 75005 Paris, France }

\author{X. Liu}
\affiliation{Sorbonne Universit\'e, CNRS, Laboratoire de Chimie Physique -- Mati\`ere et Rayonnement (LCPMR), 75005 Paris, France }

\author{D. Weder}
\affiliation{Max Born Institut f\"ur Nichtlineare Optik und Kurzzeitspektroskopie, 12489 Berlin, Germany}

\author{C. von Korff Schmising}
\affiliation{Max Born Institut f\"ur Nichtlineare Optik und Kurzzeitspektroskopie, 12489 Berlin, Germany}

\author{M. Schneider}
\affiliation{Max Born Institut f\"ur Nichtlineare Optik und Kurzzeitspektroskopie, 12489 Berlin, Germany}

\author{C. M. G\"unther}
\affiliation{Technische Universit\"at Berlin, Institut f\"ur Optik und Atomare Physik, Stra\ss e des 17. Juni 135, 10623 Berlin}
\altaffiliation{Current address: Technische Universit\"at Berlin, Zentraleinrichtung Elektronenmikroskopie (ZELMI), Stra\ss e des 17. Juni 135, 10623 Berlin}

\author{B. Mahieu}
\affiliation{Laboratoire d'Optique Appliqu\'ee (LOA), ENSTA ParisTech, CNRS, Ecole Polytechnique, Universit\'e Paris-Saclay, 828 boulevard des Mar\'echaux, 91762 Palaiseau Cedex, France}

\author{G. Malinowski}
\affiliation{Institut Jean Lamour, Universit\'e Henri Poincar\'e, Nancy, France}

\author{M. Hehn}
\affiliation{Institut Jean Lamour, Universit\'e Henri Poincar\'e, Nancy, France}

\author{D. Lacour}
\affiliation{Institut Jean Lamour, Universit\'e Henri Poincar\'e, Nancy, France}

\author{F. Capotondi}
\affiliation{FERMI, Elettra-Sincrotrone Trieste, SS 14 - km 163.5, 34149 Basovizza, Trieste, Italy }%

\author{E. Pedersoli}
\affiliation{FERMI, Elettra-Sincrotrone Trieste, SS 14 - km 163.5, 34149 Basovizza, Trieste, Italy }%

\author{I. P. Nikolov}
\affiliation{FERMI, Elettra-Sincrotrone Trieste, SS 14 - km 163.5, 34149 Basovizza, Trieste, Italy }%

\author{V. Chardonnet}
\affiliation{Sorbonne Universit\'e, CNRS, Laboratoire de Chimie Physique -- Mati\`ere et Rayonnement (LCPMR), 75005 Paris, France }

\author{E. Jal}
\affiliation{Sorbonne Universit\'e, CNRS, Laboratoire de Chimie Physique -- Mati\`ere et Rayonnement (LCPMR), 75005 Paris, France }

\author{J. L\"{u}ning}
\affiliation{Sorbonne Universit\'e, CNRS, Laboratoire de Chimie Physique -- Mati\`ere et Rayonnement (LCPMR), 75005 Paris, France }
\affiliation{Synchrotron SOLEIL, L'Orme des Merisiers, Saint-Aubin, 91192 Gif-sur-Yvette, France}

\author{B. Vodungbo}
\email{boris.vodungbo@sorbonne-universite.fr}
\affiliation{Sorbonne Universit\'e, CNRS, Laboratoire de Chimie Physique -- Mati\`ere et Rayonnement (LCPMR), 75005 Paris, France }

\date{\today} % It is always \today, today, but any date may be explicitly specified

\begin{abstract}

We use time-resolved x-ray resonant magnetic scattering (tr-XRMS) at the Co M$_{2,3}$- and Tb O$_1$-edges to study ultrafast demagnetization in an amorphous Co$_{88}$Tb$_{12}$ alloy with stripe domains. 
Combining the femtosecond temporal with nanometer spatial resolution of our experiment, we demonstrate 
that the equilibrium spin texture of the thin film remains unaltered by the optical pump-pulse on ultrashort timescales ($<$1\,ps). However, after $\simeq$ 4\,ps, we observe the onset of a significant domain wall broadening, which we attribute to a reduction of the uniaxial magnetic anisotropy of the system, due to energy transfer to the lattice. Static temperature dependent magnetometry measurements combined with analytical modeling of the magnetic structure of the thin film corroborate this interpretation.

\end{abstract}

\maketitle

\section{Introduction}

In their seminal work published in 1996, Beaurepaire \textit{et al.} demonstrated that femtosecond infrared laser pulses can induce a transient quenching of the magnetic moment of a Ni thin film on sub-ps timescales\cite{Beaurepaire1996}. This unexpected finding gave birth to a whole new research field, ``femtomagnetism'', which aims at unraveling the complex non-equilibrium phenomena at play when magnetic nanostructures and thin films are subjected to ultrashort optical excitations. However, despite more than 20 years of scientific effort, there is still no consensus regarding the underlying microscopic mechanisms and a variety of theories are currently  ``coexisting''\cite{Carpene2008,Krauss2009,Bigot2009,Koopmans2010,Battiato2010,Illg2013,Zhang2018,Dewhurst2018,Dornes2019}.  
One particularly appealing explanation for the occurence of laser-induced demagnetization is the creation of spin-polarized currents, resulting from the different mean free paths of majority and minority spin carriers\cite{Battiato2010,Battiato2012}. In fact, during the last decade, a plethora of carefully designed experiments have been performed unraveling the importance of such superdiffusive spin flow during ultrafast magnetization loss\cite{Malinowski2008,Vodungbo2012,Rudolf2012,Pfau2012,Graves2013,Wieczorek2015,Elyasi2016,Tengdineaap2018}.

\begin{figure*}
\begin{picture}(1000,200)
\put(5,-10){\includegraphics[scale=1.0]{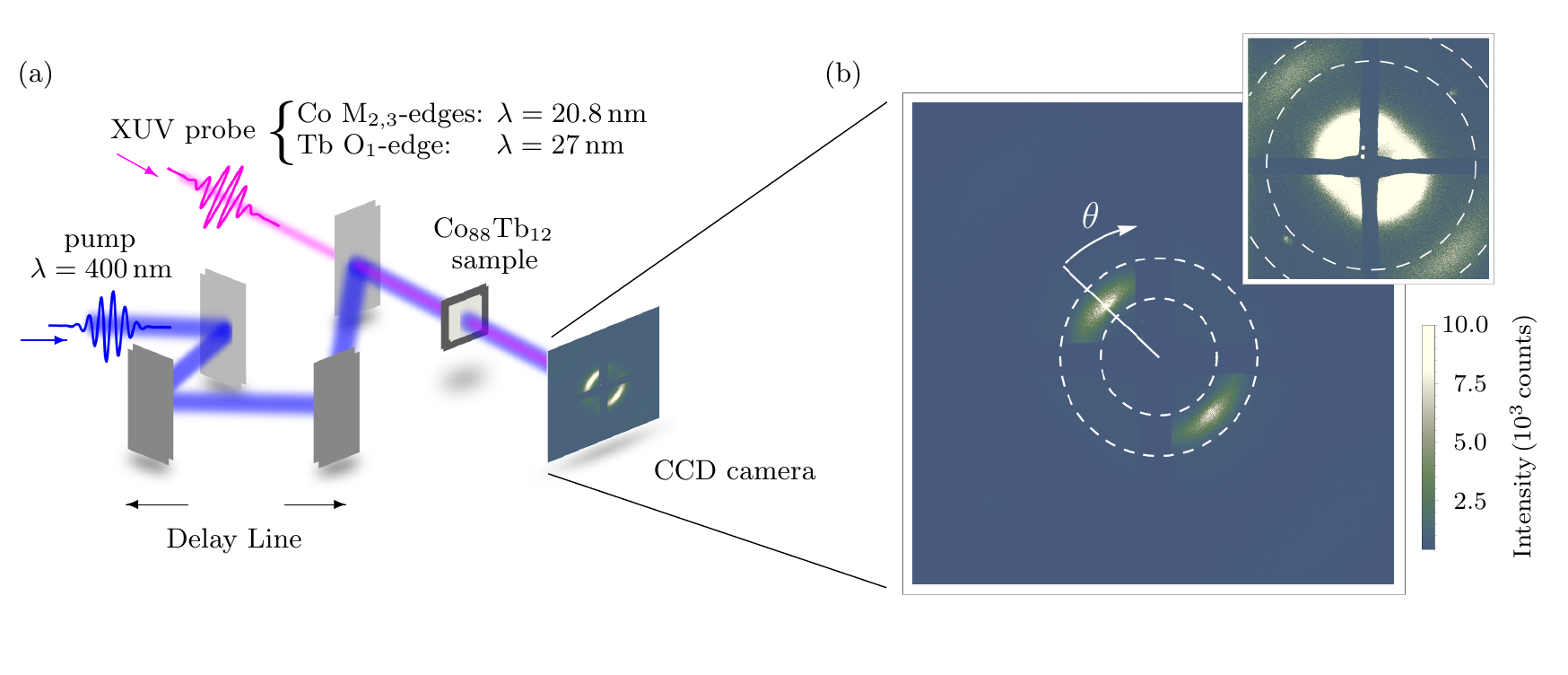}}
\end{picture}
\caption{\label{Fig:1}Schematic of the pump-probe experiment and exemplary CCD data. (a) Resonant magnetic scattering setup: the CoTb samples are pumped using a laser with $\lambda = 400\,$nm (blue beam) and varying fluences up to $\Phi = 5.5\,$mJ$\cdot$cm$^{-2}$. The system is then probed using XUV radiation (magenta beam) tuned either to the Co M$_{2,3}$-edges ($\lambda=20.8$\,nm, $E = 59.6$\,eV) or O$_{1}$-edge ($\lambda=27$\,nm, $E = 45.9$\,eV). Changing the delay $\Delta t$ between pump and probe pulses allows to record the temporal evolution of the scattering signal. (b) Magnetic diffraction pattern obtained at the Co M$_{2,3}$ edges (1024 px $\times$ 1024 px) - the first order magnetic scattering peaks can clearly be identified. The azimuthal angle $\theta$ used for integration of the signal is sketched as well (the azimuthal maximum of the peak defines $\theta=0$). Inset: The same data set is plotted using a different intensity scale (same lower bound, but saturated beyond $I_{\mathrm{max}}$\,=\,500\,counts) to highlight the third order diffraction maxima. The small peaks that can be observed at $\theta = 90^{\circ}$ result from an artificial grating structure manufactured prior to the magnetic film deposition on the back side of the sample membrane and are not relevant in the context of this study\cite{Schneider2018}.}
\end{figure*}

Magnetic thin films with strong perpendicular magnetic anisotropy (PMA) and presenting alternating domains with ``up'' and ``down'' magnetization are a beautiful testbed to gain a better understanding of such ultrafast spin transport phenomena. Indeed, they are ideally suited to perform x-ray resonant magnetic scattering (XRMS) experiments\cite{Kortright2001,Hellwig2003}, which provide intrinsic element selectivity and nanometric spatial resolution. Using appropriate pump-probe geometries, these advantages can be combined with femtosecond temporal resolution, making tr-XRMS a powerful tool to study ultrafast spin texture changes. Yet, the results gathered on different magnetic systems remain contradictory and their interpretation controversial. 
Pfau \textit{et al.} studied Co/Pt multilayers at the Co 3p$_{3/2}$ absorption resonance and found an ultrafast shift of the maximum value of the scattering intensity with a time constant of approximately 300\,fs, which they attributed to a superdiffusive spin-current-induced domain wall broadening\cite{Pfau2012}. In a similar experiment, Zusin \textit{et al.} scrutinized CoFe/Ni multilayers at the Ni L$_3$ edge and identified a pump-induced shift of the diffraction ring reaching 6\% within 1.6\,ps which they interpreted as an ultrafast domain dilation, resulting from inelastic electron-magnon scattering\cite{Zusin2020}. This contrasts with earlier work performed by Vodungbo \textit{et al.}, who studied Co/Pd multilayers at the Co M$_{2,3}$-edges using a fs high-order harmonic generation (HHG) source and reported that the  magnetic domain structure remained unaffected during the ultrafast demagnetization process\cite{Vodungbo2012}. However, they found a significantly decreased demagnetization time, which they attributed to an ultrafast angular momentum transfer between adjacent domains. Surprisingly, Moisan \textit{et al.}, studying maze patterned Co/Pt and Co/Pd thin films using a time-resolved magneto-optical Kerr (tr-MOKE) setup, concluded that hot electron spin dependent transfer between adjacent domains does not impact the ultrafast dynamics at all\cite{Moisan2014}. These examples clearly illustrate the need to conduct additional studies in order to (a) further quantify the impact of ultrashort laser excitation on magnetic thin films with complex spin texture and (b) clarify the role played by superdiffusive spin currents when changes of the magnetic structure occur on femtosecond timescales. 

In the present work, we describe tr-XRMS experiments on amorphous Co-Tb thin films with magnetic stripe domains conducted at the free-electron laser FERMI. Although several femtomagnetism studies have already been performed on ferrimagnetic Co-Tb\cite{LopezFlores2013,Bergeard2014,Alebrand2014,Ferte2017}, a material system of great technological relevance for future all optical magnetic data storage\cite{Stanciu2007,Alebrand2012,Hadri2016}, data describing the evolution of magnetic domain structures in rare earth - transition metal (RE-TM) alloys following an ultrashort optical pulse are still scarce. Recently, Fan \textit{et al.} performed a first tr-XRMS study on Co$_{88}$Tb$_{12}$ samples using a tabletop HHG source but their analysis remained limited to the first magnetic diffraction order at the N-edge of Tb (155\,eV) \cite{Fan2019}. In the present work, we complement their findings using different probe beam energies and deepen the analysis by using an experimental setup that allows to record the first and third magnetic diffraction order simultaneously. With this, we are able to explicitely monitor the pump-induced evolution of the periodic magnetic structure, \textit{i.e.}, the change of domain size and domain wall width with highest accuracy up to 120\,ps. Our results provide additional insight into the temporal evolution of the system and challenge earlier findings, as will be shown in detail in the following.  

\section{Experiments}

\subsection{Sample growth and characterization}

The samples consist of Co-rich\cite{Gottwald2012}, amorphous Co$_{88}$Tb$_{12}$ thin films with nominal thickness $h = 50$\,nm, obtained via sputter deposition on Si$_{3}$N$_4$ square membranes with surface 50 \textmu m $\times$ 50 \textmu m covered with a buffer layer consisting of 5\,nm Pt on top of 5\,nm Ta. A 3\,nm Pt capping layer was used to protect the sample from oxidation. After deposition, the samples were subjected to a demagnetization procedure using an oscillating in-plane field with decreasing amplitude\cite{Hellwig2003}. 

Polar magneto-optical Kerr effect (p-MOKE) measurements were performed to analyze the static magnetic properties of the sample in out-of-plane direction. Magnetic force microscopy (MFM) data was gathered on an Asylum Research magnetic force microscope. Temperature dependent in-plane magnetization curves were obtained using a vibrating sample magnetometry / superconducting quantum interference device (VSM/SQUID) from Quantum Design. 

\subsection{Time resolved x-ray resonant magnetic scattering (tr-XRMS)}

The pump-probe experiments were performed at the DiProI beam line of the free-electron laser (FEL) FERMI\cite{Capotondi2013}. The samples were pumped with a laser ($\lambda=400$\,nm) with 60\,fs pulse duration, then probed by 70\,fs short monochromatic circularly polarized XUV pulses delivered by the FEL with photon energies tuned either to the Co M$_{2,3}$ absorption edges at 20.8 nm (59.6 eV) or the Tb O$_1$ absorption edge at 27
nm (45.9 eV) with a beam monochromaticity $\frac{\Delta \lambda}{\lambda} \simeq \mathcal{O}(10^{-3})$ (FWHM). The delay line between the two pulses was implemented on the optical path of the pump pulse. 
A four quadrant photodiode was used to provide an accurate shot to shot measurement of the
FEL intensity, and allowed to normalize the detected scattering intensity by the incoming
photon flux ($I_0$) and to monitor the pointing stability of the beam. The scattering patterns were detected with a vacuum compatible charge coupled device (CCD) with 2048 $\times$ 2048 px$^2$ and a pixel width of 13.5 \textmu m. To increase the read-out speed, the CCD pixels have been binned 2 by 2 yielding an effective pixel size of 27 \textmu m. The spot size diameter of the probe and the pump beam were equal to 190 $\times$ 180 \textmu m$^2$ and
400 $\times$ 400 \textmu m$^2$, respectively. This focal size ensured a uniform illumination across each single membrane. To block the direct beam, a cross-shaped beam stop was placed in front of the CCD. A schematic of the experimental setup is depicted in Fig.\,\ref{Fig:1}. 

\section{Results and Discussion}

\subsection{PMA and creation of stripe domains}

\begin{figure}
\begin{picture}(210,210)
\put(-10,0){\includegraphics[scale=1.0]{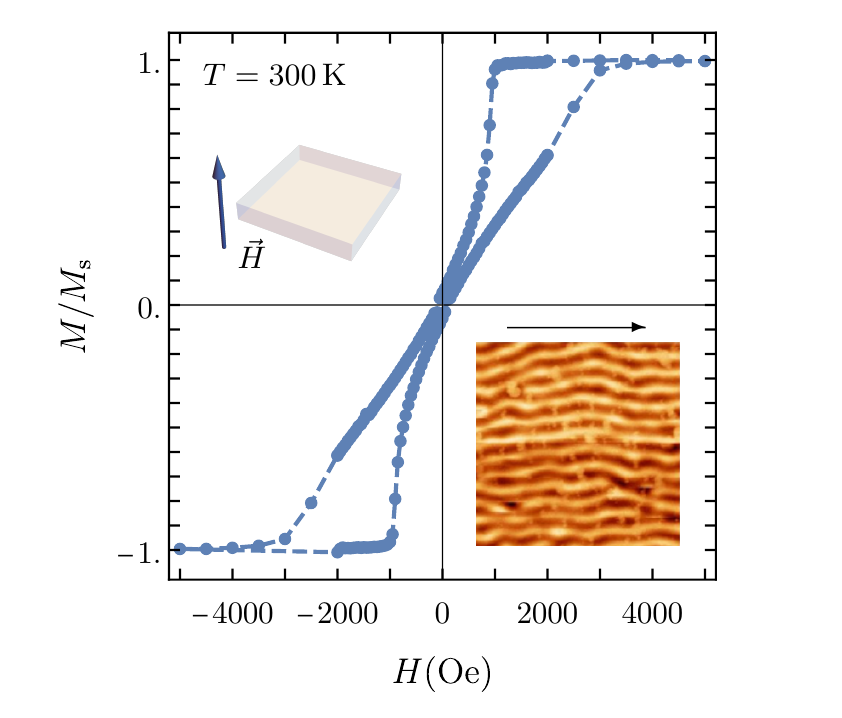}}
\end{picture}
\caption{\label{Fig:2}Room temperature p-MOKE measurements describing the out-of-plane (OP) magnetization of the sample as a function of the applied external field. Inset: MFM-phase contrast image (2$\times$2 \textmu m$^2$). The arrow on top of the scan indicates the direction of the external magnetic field applied during the demagnetization procedure.}
\end{figure}

Despite their amorphous structure, sputter deposited RE-TM thin films can exhibit strong PMA as a result of local chemical ordering\cite{Harris1992}, giving rise to domains with alternating ``up'' and ``down'' magnetization\cite{Kittel1949}. The presence of such domains in our thin films was checked using p-MOKE measurements at room temperature, yielding information about the field dependence of the average magnetization in out-of-plane direction. As shown in Fig.\,\ref{Fig:2}, the hysteresis loop presents a shape that is characteristic for magnetic stripe and maze patterned films\cite{Hellwig2003,Moisan2014}. Starting from saturation at high external fields, domains start to nucleate and grow upon reduction of $H$ ( $\simeq$ 1\,kOe). For zero external field, the magnetic moments of ``up'' and ``down'' domains compensate, resulting in an almost vanishing remanent magnetization. Reversal and further decrease of \emph{\textbf{H}} then leads to successive annihilation of domains with \emph{\textbf{M}} pointing in opposite direction, eventually yielding a saturated monodomain configuration\cite{Hellwig2003}. 

Information about the spatial distribution of magnetic domains was obtained using MFM measurements. As shown in the inset of Fig.\,\ref{Fig:2}, a stripe domain structure resulting from the demagnetization procedure can clearly be evidenced and confirms our MOKE-based analysis. The periodicity determined from a fast Fourier transformation of these scans is $ \Lambda \simeq$ 220\,$\pm$\,10 nm. This corresponds to an average domain size $ w = \frac{\Lambda}{2} \simeq$ 110\,$\pm$ 5 nm. 

\begin{figure}
\centering
\begin{picture}(210,210)
\put(-10,0){\includegraphics[scale=1.0]{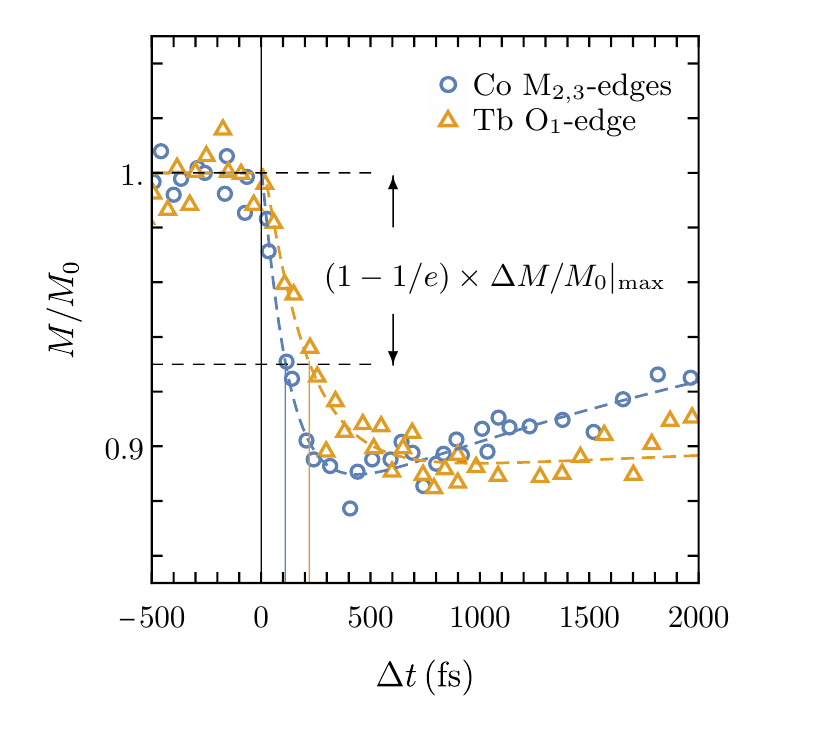}}
\end{picture}
\caption{\label{Fig:3}Temporal evolution of the domain magnetization $M$ (normalized to the average unpumped value $M_0$) measured at the Co M$_{2,3}$-edges (blue open circles) and Tb O$_1$-edge (orange triangles) using a pump fluence $\Phi =$ 5.5\,mJ$\cdot$cm$^{-2}$ and fits to the 3TM (Appendix A). The time required to reach $1-1/e = 63.2\%$ of the maximum demagnetization amplitude $\Delta M/M_0|_{\mathrm{max}}$ is shown as well: $\tilde{\tau}(\mathrm{Co})=107$\,fs and $\tilde{\tau}(\mathrm{Tb})=216$\,fs.}
\end{figure} 

\subsection{Ultrafast demagnetization}

The laser excitation-induced evolution of the magnetization in CoTb was analyzed with tr-XRMS using the stripe domain samples described in the last paragraphs. To study the Co and Tb sublattices individually, the probe beam energy was set either to the M$_{2,3}$-edges of Co, or to the O$_{1}$-edge of Tb. For every delay value $\Delta t$, CCD images were obtained by averaging over 50 pulses (5\,s aquisition time with the FEL operated at 10\,Hz) and then background subtracted. An exemplary intensity map, \textit{i.e.}, CCD data, is shown in Fig.\,\ref{Fig:1}(b). The appearance of well defined maxima results from x-ray magnetic circular dichroism (XMCD) and reflects the periodic structure of the sample which acts as a magnetic grating for the incoming soft x-ray probe pulses. For symmetry reasons, \textit{i.e.}, identical width of up and down domains, only uneven diffraction orders can be observed. The four peaks seen on the camera (Fig.\,\ref{Fig:1}) thus correspond to the positive and negative first ($n=1$) and third magnetic diffraction order ($n=3$).
As shown in earlier work for magnetic systems with perfect square wave magnetization pattern, their intensity is proportional to the square of the scalar product between the domain magnetization vector \emph{\textbf{M}} and the propagation vector \emph{\textbf{k}} of the x-ray beam\cite{Kortright2001}. Note that, due to non-parallel in-plane alignment of the stripes in the probed zone, the peaks are smeared along the arc of a circle (Fig.\,\ref{Fig:1}), which, for the extreme case of a magnetic maze pattern, would give rise to homogeneous diffraction rings. Applying these considerations to our specific experimental setup results in $ M(\Delta t) \propto \sqrt{I_1(\Delta t)}$, which links the local domain magnetization $M(\Delta t)$ with $I_1(\Delta t)$, the total first order peak intensity after azimuthal integration. We stress that in our approach, we implicitely neglected possible non-magnetic contributions that might result from charge heterogeneities on nanometer lengthscales\cite{Kortright2001}. This was motivated by the absence of any significant signal measured at $\theta = 90^{\circ}$, where the magnetic contrast (essentially resulting from defects along the stripe patterns) ought to be negligible.

Figure\,\ref{Fig:3} shows the normalized magnetization $M(\Delta t)/M_0$ for Co and Tb atoms obtained with the maximum pump fluence employed in the present work ($\Phi=5.5$\,mJ$\cdot$cm$^{-2}$). Both curves exhibit a rather similar behavior: The intensity drops
rapidly by approximately 10\% during the first hundreds of femtoseconds, and then slowly increases on longer timescales. While the degree of demagnetization is comparable for both elements, the characteristic times associated with the initial quenching and subsequent magnetization recovery clearly differ. To gain further insight into these processes, we fitted the data to the well established three temperature model\cite{Beaurepaire1996} (see Appendix\,A for details). 

\begin{table}
\caption{Impact of the pump fluence $\Phi$ on the maximum demagnetization $\frac{M - M_0}{M_0}$ and characteristic demagnetization time constants at the Co M$_{2,3}$-edges obtained by fitting the data to the 3TM (Appendix\,A).}
\begin{center}
\begin{tabular}{ccc}
$\Phi$ (mJ$\cdot$cm$^{-2}$) & $\frac{M_0 - M}{M_0}|_{\mathrm{max}}$ & $\tau_M$ (fs) \\
\hline
2.6    & 0.06  & 110 $\pm$ 30 \\
3.4    & 0.08  & 105 $\pm$ 30 \\
5.5    & 0.11 & 110 $\pm$ 30 \\
\end{tabular}
\end{center}
\end{table}

The fits, presented in Fig.\,\ref{Fig:3}, provide a quantitative description of the demagnetization process and allow to draw several conclusions: First, within experimental resolution ($\simeq 50$\,fs), we do not observe any delay between the demagnetization onset of Co and Tb atoms\cite{Lai2019}.
Second, the two magnetic sublattices exhibit very similar demagnetization amplitudes. Finally, the demagnetization process in Tb is found to be slower than in Co, which is in agreement with earlier reports on RE-TM alloys\cite{Radu2011,LopezFlores2013,Bergeard2014}. Surprisingly, from a quantitative perspective, our results show rather poor agreement with literature data. At the Co edge, we obtain $\tau_M(\mathrm{Co}) = 110 \pm 30$\,fs (with almost no impact of the employed pump fluence, as shown in Table\,I), which is slightly smaller than what has been reported so far for pure Co thin films\cite{Koopmans2010}, as well as for ferro-\cite{Boeglin2010,Moisan2014} and ferrimagnetic\cite{LopezFlores2013,Jal2019,Radu2015} Co-based alloys. At the Tb edge, where our fits yield $\tau_M(\mathrm{Tb}) = 220 \pm 30$\,fs, the discrepancy is even more pronounced. Indeed, this value is 2-3 times smaller than what has been obtained on CoTb thin films with almost identical composition\cite{LopezFlores2013,Fan2019}.

It was put forward that decreased values of $\tau_M$ might be an indicator for superdiffusive spin transport in stripe or maze domain structured films. Observations made by Vodungbo \textit{et al.} in CoPd alloys with large PMA\cite{Vodungbo2012} revealed surprisingly small demagnetization time constants ($\tau_M \simeq$ 100\,fs), that remained independent of the fluence, and which they attributed to a spin-polarized transport of hot electrons between neighboring domains. However, in the present case, we believe that $\tau_M$ values alone can hardly serve as conclusive parameters to evidence the presence of ultrafast spin currents, considering (i) the rather large spread of literature values, (ii) the different degrees of quenching obtained as well as (iii) the variety of experimental techniques (XMCD at different edges, MOKE) employed in previous studies to characterize the pump-induced magnetization drop. 

A more reliable strategy to observe superdiffusive spin transfer between domains would consist in unveiling possible ultrafast changes of the domain wall structure, where an enhanced demagnetization due to an accumulation of minority electrons is expected to take place\cite{Pfau2012,Sant2017}. As will be shown in the next section, our experiments, where the first and third order magnetic diffraction peaks are recorded simultaneously, allow to directly monitor the impact of the pump pulse on the magnetic structure of the CoTb thin films, thereby providing more reliable evidence for the presence or absence of superdiffusive interdomain spin transport.

\begin{figure}[h!]
\begin{picture}(210,390)
\put(-10,0){\includegraphics[scale=1.0]{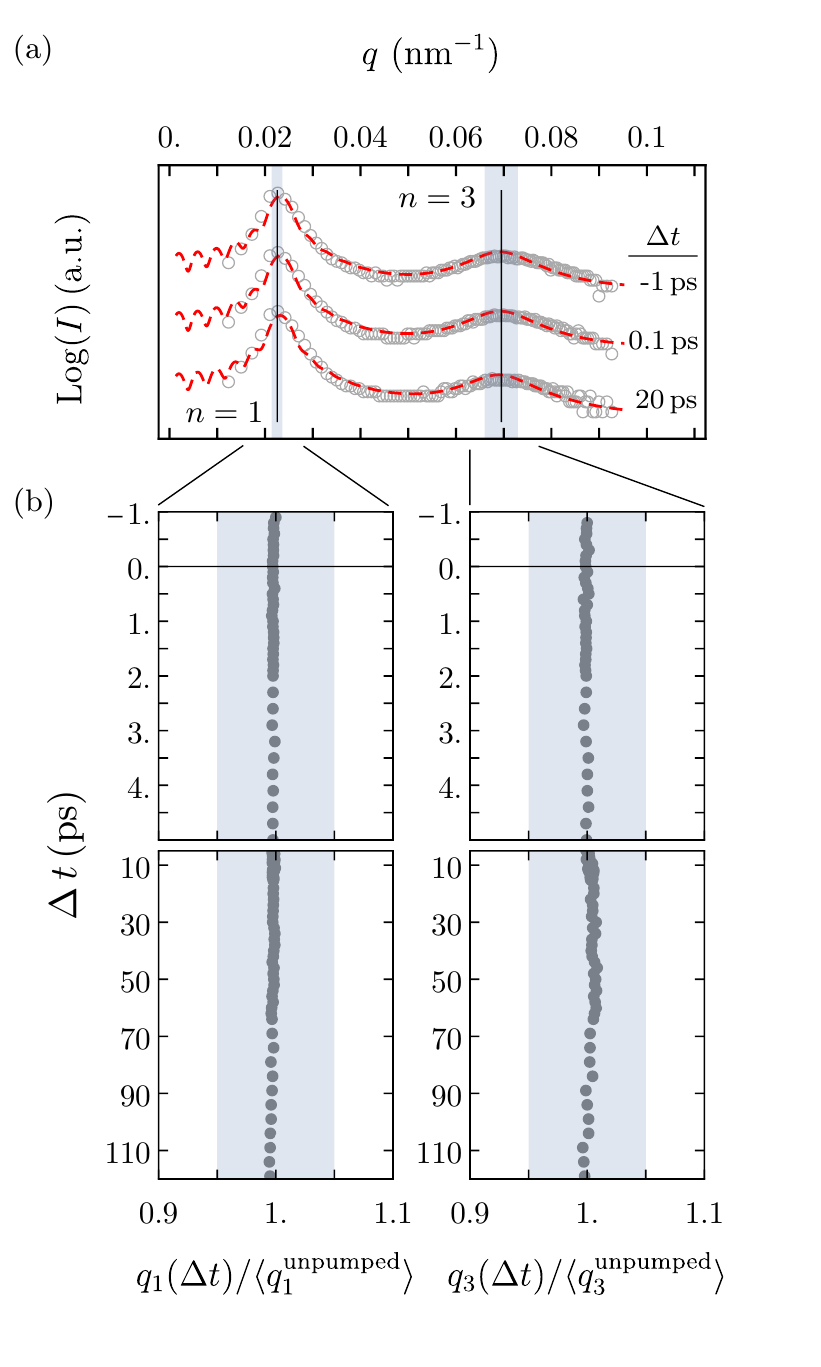}}
\end{picture}

\caption{\label{Fig:4}Azimuthally integrated intensity in reciprocal space and temporal evolution of the first and third order peaks with time. (a) The CCD intensity is translated into reciprocal space (Appendix B) and plotted as a function of $q$ on a logarithmic scale after azimuthal integration over $ -\pi/24 <\theta< \pi/24$ (see Fig.\,\ref{Fig:1}, where the white line indicates $\theta = 0$). The results are shown for three different delays $\Delta t$ (gray circles), the curves have been shifted vertically for better readability of the data. The positions of the first order ($n=1$) and third order maxima ($n=3$) are highlighted. Fits to the Hellwig model are shown as red dashed lines. (b) Temporal evolution of the first order (left) and third order peak maxima (right), obtained from a Gaussian fit and normalized by the unpumped value. The blue background corresponds to a 5\% variation of the peak positions in reciprocal space.}  
\end{figure}

\subsection{Pump-pulse-induced changes of the magnetic structure}

In the previous section, we have used the first order magnetic diffraction peak intensity to gain information about the average domain magnetization. In fact, additional spatial information concerning the spin texture of the thin film, \textit{i.e.}, average domain sizes $w(\Delta t)$ and domain wall widths $d(\Delta t)$, is encoded in the time-dependent diffraction pattern. To calculate $w(\Delta t)$, we first translated the CCD data into $q$-space (Appendix\,B). After subsequent azimuthal integration, the average domain size was obtained from $I(q)$ via $w(\Delta t) = \frac{\pi n}{q_n(\Delta t)}$, $q_n(\Delta t)$ being the location of the $n$-th maximum for a given delay $\Delta t$. All results presented in the following were obtained from data gathered at the Co-edge, using the maximum pump fluence $\Phi = 5.5\,$mJ$\cdot$cm$^{-2}$.

Figure\,\ref{Fig:4}(a) shows $I(q)$ curves gained after azimuthal integration for different delays $\Delta t$, Fig\,\ref{Fig:4}(b) the temporal evolution of $q_1(\Delta t)$ and $q_3(\Delta t)$, the first order and third order maxima. For better readability, the $q_n$ values were normalized using the unpumped references 
%$\langle q_1^{\mathrm{unpumped}}\rangle = 0.0225$\,nm$^{-1}$ and $\langle q_3^{\mathrm{unpumped}}\rangle = 0.0698$\,nm$^{-1}$. 
$\langle q_1^{\mathrm{unpumped}}\rangle = 0.023$\,nm$^{-1}$ and $\langle q_3^{\mathrm{unpumped}}\rangle = 0.070$\,nm$^{-1}$. 
This corresponds to an average domain size of $136 \pm 1$\,nm, which is slightly larger, but still in reasonable agreement with our MFM measurements and close to the values obtained by Fan \textit{et al.} on a Co$_{88}$Tb$_{12}$ sample with identical magnetic layer thickness\cite{Fan2019}. Careful analysis of the data reveals that there is no ultrafast $q_n$ variation for the fluences employed in our study: Neither for sub-picosecond delays, as reported in Co/Pt multilayers ($\Delta q/q_0=4\%$ after 300\,fs)\cite{Pfau2012} nor on slightly longer timescales, as recently claimed for CoFe/Ni thin films ($\Delta q/q_0=6\%$ after 1.6\,ps)\cite{Zusin2020}, do we find any significant shift of the magnetic diffraction peaks. 
%We conclude that, within experimental accuracy ($\simeq 1\%$) and for the pump-fluences employed in our study, the optical excitation does not induce measurable changes of the domain size.

\begin{figure}[h!]
\begin{picture}(210,200)
\put(-15,0){\includegraphics[scale=1.0]{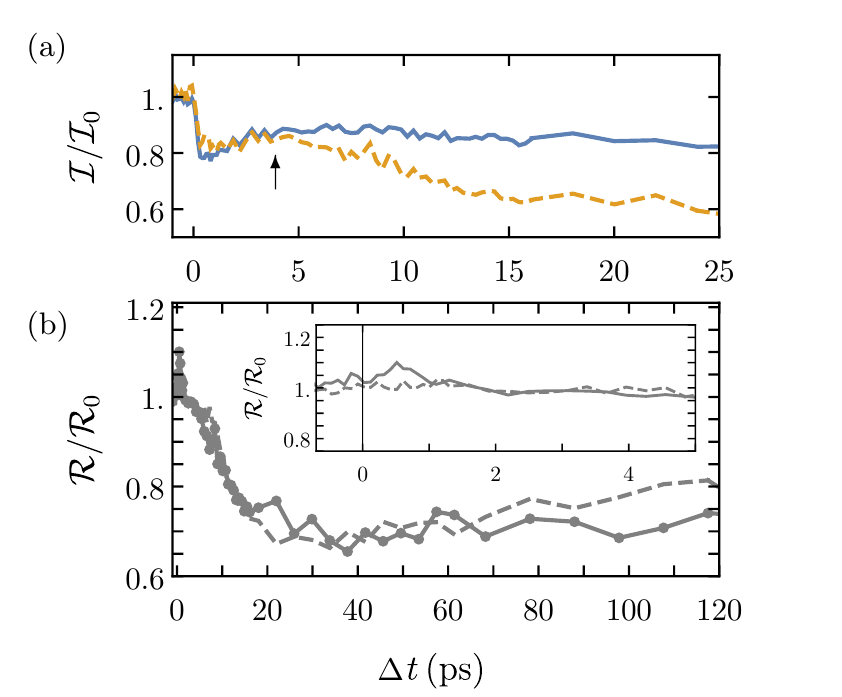}}
\end{picture}
\caption{\label{Fig:5}Relative temporal evolution of (a) the integrated first and third order diffraction peaks (blue, solid line and orange, dashed line, respectively) and (b) their ratio $\mathcal{R}/\mathcal{R}_0$ shown for two different, consecutive delay scans, illustrating the good reproducibility of the measurements. The black arrow shown in (a) indicates the onset of the domain wall broadening.}  
\end{figure}

In addition to the temporal evolution of the domain periodicity, the simultaneous recording and analysis of more than one diffraction peak also allows to assess the domain wall width, as has been proposed in earlier studies\cite{Vodungbo2012}. We show in Appendix\,C that in a simple linear chain approximation, $\mathcal{R} = I_3/I_1$, the ratio of integrated first and third order peak intensities depends exclusively on $\eta = d/w$, the ratio of linear domain wall and average domain size. Considering the constant domain size evidenced at the beginning of this section, any change of $\mathcal{R}$ must necessarily result from a variation of the domain wall width. Figure\,\ref{Fig:5}(a) shows the time dependency of the integrated first and third order intensities, normalized to their unpumped values ($\Delta t < 0$). As can be seen, both exhibit the same temporal evolution during the first several ps. This becomes even clearer when plotting the normalized ratio $\mathcal{R}(\Delta t)/\mathcal{R}(0)$, which corresponds to the ratio of normalized third and first order integrated peaks, shown for two independent measurements in Fig.\,\ref{Fig:5}(b). As presented in the inset of Fig.\ref{Fig:5}(a), no change of the DW width, other than random fluctuations (up to 10\% for a single measurement), can be identified for delays below 4\,ps. This is another central finding of the present study: cascades of spin polarized electrons, which are believed to create an ultrafast enhanced demagnetization at the domain boundaries, \textit{i.e.}, a sub-ps broadening of the domain walls, do not seem to play a significant role in the present system. %at least, for fluences up to $\Phi = 8 mJ/cm^{-2}$.
However, what becomes apparent in Fig.\ref{Fig:5}(a), and even more manifest in Fig.\ref{Fig:5}(b), is a pronounced decrease of the ratio of third and first order intensities for $\Delta t \gtrsim 4$\,ps. Indeed $\mathcal{R}(\Delta t)/\mathcal{R}(0)$ drops to 0.73 between $ 4\,\mathrm{ps} \lesssim t \lesssim 20$\,ps and then slowly increases again for $\Delta t > 40$\,ps, hinting at pronounced modifications of the DW width on picosecond timescales. 

\begin{figure}
\begin{picture}(210,220)
\put(-12,0){\includegraphics[scale=1.0]{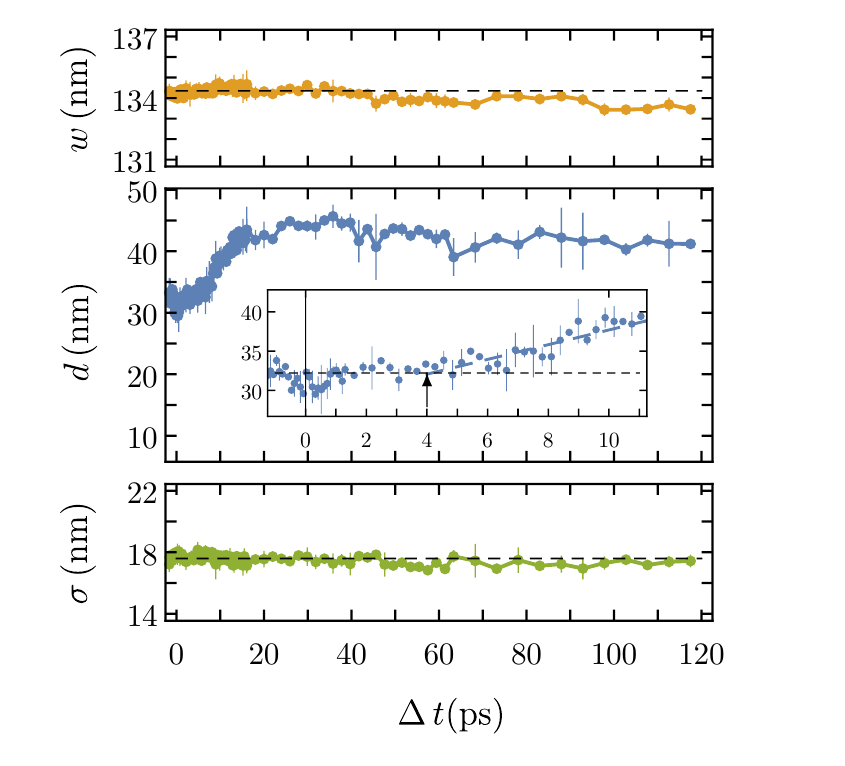}}
\end{picture}
\caption{\label{Fig:6}Hellwig's model fitted to the present XRMS data at the Co $M_{2,3}$-edges. Evolution of the fit parameters with delay $\Delta t$, from top to bottom: domain size $w$ (orange), domain wall width $d$ (blue) and width of the domain size distribution $\sigma$ (green). While $w$ and $\sigma$ remain constant over time, a clear increase of the domain wall size can be seen for $\Delta t > 4$\,ps (black arrow).}
\end{figure}

In principle, the aforementioned analysis can be used to obtain a quantitative description of the temporal evolution of the DW size (Appendix C). However, this approach is limited due to the presence of the beam-stop, which hampers data integration at low $q$. In addition, we stress that the underlying model is unable to account for a finite correlation length of the stripe patterns and a distribution of domain sizes, typically encountered in our experiments (see inset of Fig.\,\ref{Fig:2}). To avoid these shortcomings, we fitted our $q$-space data using the approach proposed by Hellwig \textit{et al.}, specifically dedicated to the description of XRMS experiments on stripe domains\cite{Hellwig2003} (see Fig.\,\ref{Fig:4}(a)). Hellwig's model relies on the use of 5 adjustable parameters: $w$, $d$, the width of the domain size distribution $\sigma$ and $\mathcal{M}_z$, which is proportional to the $z$ component of the magnetization, as well as $N$, the number of stripes that scatter coherently. To facilitate the fit procedure, $N$ was kept fixed ($N=6$). Figure\,\ref{Fig:6} shows the evolution of $w$, $d$ and $\sigma$ as a function of the pump-probe delay $\Delta t$. $\mathcal{M}_z$ is omitted here: as expected, it reproduces the initial magnetization drop on sub-ps timescales. No additional and physically relevant features were identified with increasing $\Delta t$, except for a slight linear decrease of roughly 0.2\%/ps, presumably induced by absorption changes due to carbon deposition. The unpumped domain size ($w_0 = 134.3\pm 0.3$\,nm) agrees well with our earlier determination of the first and third order peak values. It does not change during the first 40\,ps and then decreases slightly ($\simeq 1$\,nm) during the following 80\,ps (note that the width of the domain size distribution is not affected by the optical pulse). Finally, the values of $d$ confirm what we have put forward earlier in this section: the domain wall width remains constant during the first ps ($ d = 32 \pm 2$\,nm). It then starts increasing significantly after 4\,ps until saturating at $45 \pm 2$\,nm after roughly 20\,ps, which represents a relative change of 41\% of its initial value. For $\Delta t>40$\,ps the domain wall width finally decreases again at a slow rate of $\simeq 0.06$\,nm\,/\,ps.  

\subsection{Lattice heating and PMA decrease}

\subsubsection{Exchange stiffness and anisotropy changes}

What causes this pronounced change of the magnetic structure? For the case of a single isolated 180$^{\circ}$ Bloch wall, an analytical expression linking $d$ to microscopic magnetic parameters is readily derived by taking into consideration the competing anisotropy and exchange contributions to the total energy \cite{OHandleyBOOK}. Minimization of the latter with respect to $d$ results in the well known expression $d = \pi \sqrt{\frac{A}{K_u}}$. Thus, pump pulse-induced changes of the exchange stiffness $A$ (which depends on the exchange integral, 
the total spin and atomic distances, as shown in Appendix\,D) as well as changes of the anisotropy constant $K_{\mathrm{u}}$ can affect the size of the domain walls. While a variation of $A$, that stems from a modification of the exchange interaction, is expected to take place on sub-fs timescales, changes linked to interatomic distances seem possible on longer timescales. In fact, lattice constant variations can result from energy flow into the phonon bath, requiring several ps for full equilibration in metals \cite{Pudell2018,Maldonado2020}. In the present case though, thermal expansion of the lattice is expected to increase interatomic distances. This would lead to a decreased exchange stiffness and, in consequence, to a reduction of domain wall sizes, in clear contradiction with our results. Note that the rather moderate laser pulse induced heating of the sample $\Delta T = \frac{R \Phi}{C_v h} \simeq 145$\,K (calculated from the reflectivity $R = 0.7$ of the thin film, heat capacity $C_{\mathrm{v}}(\mathrm{Co}_{88}\mathrm{Tb}_{12}) \simeq C_{\mathrm{v}}(\mathrm{Co}) = 3.3 \cdot 10^6$\,J/m$^3$K, assuming the energy to be distributed homogeneously over the entire film thickness) ensures that the final temperature of the system remains very far from the Curie temperature of the alloy\cite{Witter1991}. If we assume that a scaling relation similar to the one linking $A$ and $T$ in pure Co\cite{Fesenko2016} holds in our Co-rich alloy, we can expect the present temperature induced change of $A$ to be negligible. In the following, we will thus exclusively focus on $K_{\mathrm{u}}$ as the main parameter driving the observed domain wall size increase.   

\subsubsection{T-dependent magnetization measurements\\ and analytical modeling}

\begin{figure}
\begin{picture}(210,300)
\put(-15,0){\includegraphics[scale=1.0]{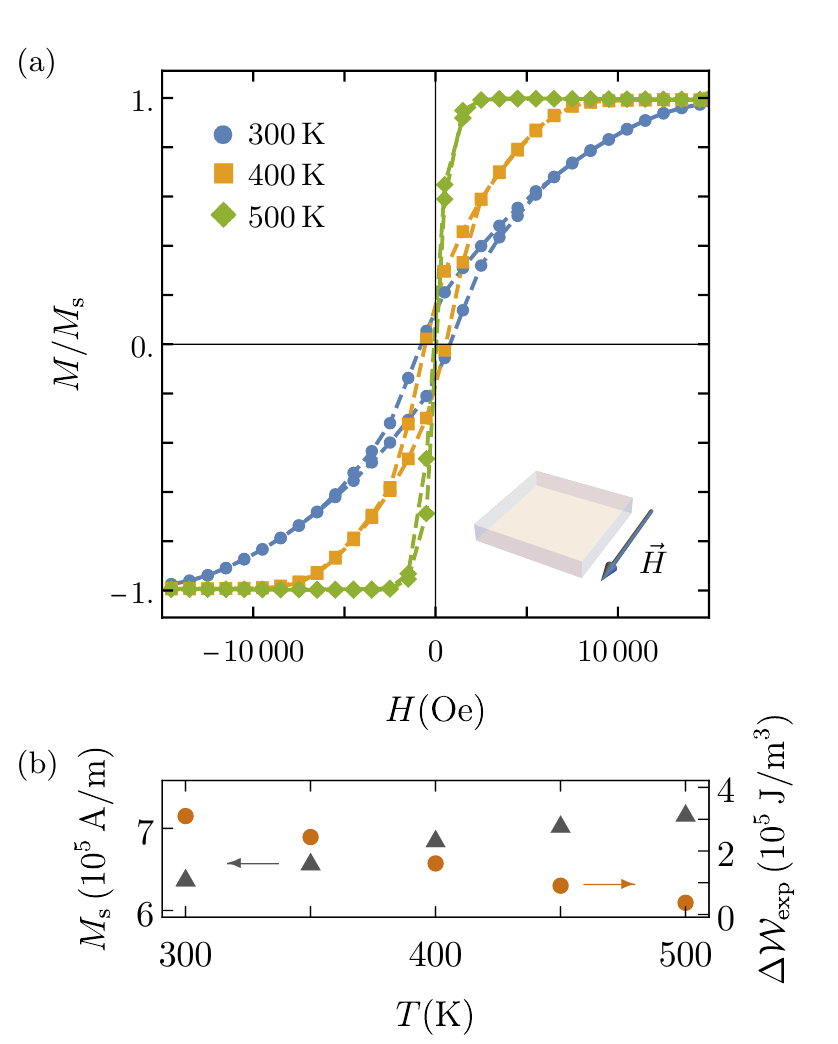}}
\end{picture}
\caption{\label{Fig:7}Temperature-dependent static VSM-SQUID measurements in IP configuration. (a) Normalized magnetization curves at 300\,K (blue dots), 400\,K (orange squares) and 500\,K (green diamonds). A small, finite hysteresis is observed for $T<500$\,K. Note that the remanent magnetization at room temperature $M_{\mathrm{r}}/M_{\mathrm{s}} \simeq 0.12 \pm 0.01$ agrees nicely with the prediction from Virot \textit{et al.}\cite{Virot2012}: $M_{\mathrm{r}}/M_{\mathrm{s}} = \frac{2 d}{\pi w} = 0.10\,\pm\,0.01$. (b) Evolution of the saturation magnetization $M_{\mathrm{s}}$ (gray triangles) and $\Delta \mathcal{W}_{\mathrm{exp}} = \mu_0\int_0^{M_{\mathrm{s}}} HdM$ (orange dots) with temperature.}
\end{figure}

Static, temperature-dependent magnetization measurements were used to compute $K_{\mathrm{u}}(T)$ and establish the link between laser induced heating of the sample and the magnetic structure changes, as will be shown in this section. 
$M(H)$ curves were determined \textit{via} SQUID-VSM measurements in in-plane geometry, with $T$ varying between 300\,K and 600\,K, as presented in Fig.\,\ref{Fig:7}(a). Qualitatively, the decreasing field values $H_{\mathrm{s}}$ at which saturation is reached already hint at the change of $K_{\mathrm{u}}$ and show that higher temperatures result in a smaller anisotropy of the system. Note that the measured values of $M_s$, presented in Fig.\,\ref{Fig:7}(b) increase with temperature, as expected for a TM-rich RE-TM alloy\cite{Gottwald2012}. 

\begin{figure}
\begin{picture}(210,205)
\put(-20,0){\includegraphics[scale=1.0]{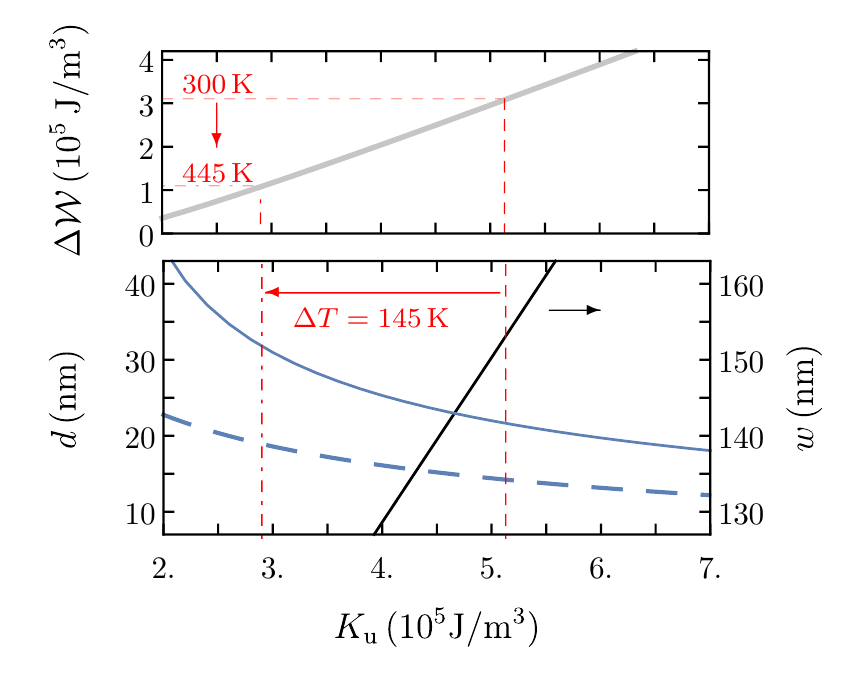}}
\end{picture}
\caption{\label{Fig:8}Virot \textit{et al.}'s sine wave model. $K_u - e_{\mathrm{tot}}(K_u) = \Delta \mathcal{W}(K_u)$, obtained by minimizing $e_{\mathrm{tot}}(K_u)$, and plotted as a function of $K_u$ (gray solid line). The experimental values $\Delta \mathcal{W}_{\mathrm{exp}}$ are shown as well (300\,K: red dash-dotted line, 445\,K: red dashed line). Domain size (black line), domain wall size in a stripe patterned thin film (blue solid line) and domain wall size of an isolated Bloch wall (blue dashed line).}
\end{figure}

For further quantitative analysis, we describe the behavior of the magnetization of the thin film upon IP field increase as a coherent hard axis magnetization process with effective anisotropy $K_{\mathrm{eff}}$\cite{OHandleyBOOK}. The latter is linked to the work $\Delta \mathcal{W}$ performed on the CoTb layer to align the magnetization in plane \textit{via} $K_{\mathrm{eff}} = \Delta \mathcal{W} = \mu_0\int_0^{M_s} H dM$, which is easily calculated from the SQUID-VSM data. The effective anisotropy in turn is obtained from $ K_{\mathrm{eff}} = K_u - e_{\mathrm{tot}}$, where $e_{\mathrm{tot}}$ represents the equilibrium energy of the stripe configuration (in OP configuration) without external field. We calculated the latter using the analytical model proposed by Virot \textit{et al.} \cite{Virot2012}. As shown in Appendix\,D, $e_{\mathrm{tot}}$ consists of a sum of  magnetostatic, anisotropy and exchange contributions and is obtained by numerical minimization with respect to $w$ and $d$. This eventually results in an implicit equation for $K_{\mathrm{u}}(\Delta \mathcal{W})$. Thus, knowledge of temperature-dependent experimental $\Delta \mathcal{W}_{\mathrm{exp}}$ values (Fig.\,\ref{Fig:7}(b)) yield $K_{\mathrm{u}}(T)$ and, consequently, the equilibrium domain wall width $d(K_{\mathrm{u}}(T))$ and domain size $w(K_{\mathrm{u}}(T))$. These quantities are plotted in Fig.\,\ref{Fig:8}, which summarizes our modeling results. 

As shown in Fig.\,\ref{Fig:7}(b), $\Delta \mathcal{W}_{\mathrm{exp}} = 3.1\cdot 10^5$J/m$^3$ at 300\,K, from which we calculate $K_{\mathrm{u}} = 5.1 \cdot 10^5$J/m$^3$ (Fig.\,\ref{Fig:8}). This is smaller than available literature data: El Hadri \textit{et al.} provide a detailed account of uniaxial anisotropy values in amorphous Co$_{x}$Tb$_{1-x}$ thin films\cite{ElHadri2016} and find $K_{\mathrm{u}} \simeq 8.1 \cdot 10^5$J/m$^3$ for $x=0.88$. But their work also illustrates how sensitive $K_{\mathrm{u}}$ reacts to composition changes. Note that the value of $M_s$ deduced for the present samples is close to the one reported in\cite{ElHadri2016} for $x=0.895$, for which $K_{\mathrm{u}} \simeq 5.9 \cdot 10^5$J/m$^3$, hinting at a slight departure from the nominal concentration in our sample. This is further confirmed by an analysis of the domain size. Indeed, our data yield $w = 151$\,nm, which is rather close to our experimental result (within $\simeq 10$\%). Interestingly, using the aforementioned Co$_{88}$Tb$_{12}$ literature value as an input would result in much too large domain sizes (beyond 200\,nm), which has further been cross-checked with the analytical approach by Kooy and Enz\cite{Kooy1960,ElHadri2016} yielding $w = 134$\,nm. Finally, $K_{\mathrm{u}} $ also allows to calculate the domain wall width. For the case of an isolated wall, we obtain $d = \pi \sqrt{\frac{A}{K_{\mathrm{u}}}} = 14$\,nm, which is less than half the experimental value and highlights the need to fully model the stripe structure of the thin film. In contrast, our approach yields $d = 21$\,nm at room temperature. While this is much closer to the experimental data, it still underestimates the measured width by almost 35\%. However, it must be kept in mind that the present model has no adjustable fit parameter and relies on a variety of approximations that deserve to be critically discussed: first, several values used as an input to our 
model are prone to large errors. This especially holds true for the exchange stiffness, which directly impacts the energy of the domain wall (Appendix D). Note that literature exchange stiffness values in CoTb alloys show a rather large spread\cite{ElHadri2016,Turner2011} $ 0.6\cdot 10^{-11} \mathrm{A/m}< A < 1.4\cdot 10^{-11} \mathrm{A/m}$. Second, one must bear in mind that the value of $d$, as determined in the present work, is only identical with the domain wall size in a thin film with ideal stripe pattern, \textit{i.e.} if the system is homogeneous along $z$. This can become an issue in magnetic layers where the projection of $M$ along $z$ changes as a function of the thin film depth. A stress induced anisotropy reduction at the interface for example (further enhanced by temperature gradients and differences in thermal expansion coefficients) could give rise to flux closure caps, inducing an ``artificial'' increase of $d$. In addition, we emphasize that the determination of $d$ from the XRMS data is based on the assumption of a simple linear wall profile, which is certainly a rather crude approximation. Finally, static heating of the thin film might lead to a working temperature exceeding room temperature by several tens of K, yielding larger domain wall widths as well as smaller domains. 

Keeping these limitations in mind, we can eventually attempt to link the deposited energy in the lattice and the resulting temperature rise to the anisotropy and domain wall size change. As shown in Fig.\,\ref{Fig:8}, a temperature increase of 145\,K, as calculated in an earlier part of this paper, results in an anisotropy reduction $\Delta K_{\mathrm{u}} = -2.2\cdot 10^5$\,J/m$^{3}$ and a domain wall size change of $\Delta d = 10$\,nm. While it compares well with our exprimental data, this result has to be considered with care though, considering the highly non-linear evolution of $d$ with $K_{\mathrm{u}}$. An identical anisotropy change from $4\cdot 10^{5}$\,J/m$^{-3}$ to $1.8\cdot 10^{5}$\,J/m$^{-3}$ for example would give rise to an absolute change more than twice as large. This problem can be circumvented by taking a closer look at the functional dependency of $d(K_{\mathrm{u}})$ and by focusing on relative, instead of absolute domain width changes. The isolated wall, as well as the stripe pattern wall widths exhibit similar power law behaviors $d \sim K_{\mathrm{u}}^{\alpha}$. The exponent $\alpha$ equals $-0.5$ for the simple Bloch wall and we find $\alpha = -0.7$ using Virot's model. This allows to write ratios of domain wall sizes as a simple function of the ratio of $K_{\mathrm{u}}$ values and to avoid the use of parameters prone to large experimental uncertainties. Using $d_{\mathrm{max}} = d_{\mathrm{i}} (K_{\mathrm{u,min}}/K_{\mathrm{u,i}})^{\alpha}$, 
where $d_{\mathrm{max}}$ is the maximum domain wall width obtained for the maximum laser induced anisotropy decrease, we take $K_{\mathrm{u,i}} = 5.1\cdot10^5$\,J/m$^3$ and the domain wall values obtained from our XRMS experiment to calculate an anisotropy decrease $\Delta K_{\mathrm{u}} = -2.5 \pm 0.3 \cdot 10^5$\,J/m$^3$ and $\Delta K_{\mathrm{u}} = -2.0 \pm 0.2 \cdot 10^5$\,J/m$^3$ for the two different wall types. This agrees nicely with our data and provides conclusive evidence for an anisotropy mediated domain wall increase.

\subsubsection{Characteristic timescales}

We close this section with some brief remarks concerning characteristic timescales linked to PMA decrease observed in our experiments. It is interesting to compare the time lapse before onset of the domain wall broadening with literature data. Zusin \textit{et al.}\cite{Zusin2020} report on a domain wall size increase setting in approximately 10\,ps after the pump pulse and which they trace back to delayed thermal diffusion through the sample thickness and a change of $K_{\mathrm{u}}$. Based on the 3TM, they provide an estimation of the time constant for heat transfer between the electronic and spin system equal to roughly 5\,ps. However, we believe that this characteristic timescale does not help in explaining the present process. In fact, the DW width changes as a consequence of a reduction of anisotropy. It is thus not an increase of the spin bath temperature which triggers the observed broadening (a strain induced modification of $K_{\mathrm{u}}$ would for example result in a similar DW width change, without requiring any temperature change). At first sight, it appears thus surprising to observe such long delays after the pump pulse absorption considering that lattice heating can be achieved on much shorter timescales\cite{Beaurepaire1996}. A possible explanation could be provided by finite spin reorientation times within the domain wall itself. Indeed, this process will not take place instantaneously, but might require several ps, considering the typical propagation speed of spin waves. 
A second relevant quantity that can be extracted from our analysis is the time required for the domain wall increase to saturate ($\Delta t \simeq 20$\,ps), which corresponds to energy transfer to the lattice degrees of freedom and eventual equilibration along $z$. This process will be accompanied by the creation of a strain wave propagating in a direction perpendicular to the thin film surface. Using the speed of sound of Co as an approximation for the present system\cite{Samsonov1968}, we calculate $\Delta t = 2h/v \simeq 21$\,ps, for the back and forth propagation of the latter through the entire film.
Finally, we stress again that no domain dilation has been observed in our experiments, in contrast to what has been reported in recent studies\cite{Fan2019,Zusin2020}. As already mentioned, Zusin \textit{et al.} interpret their observation of a domain wall broadening for $\Delta t > 10$\,ps as a result of a lattice heating induced anisotropy reduction. Surprisingly, they simultaneously observe an increase of the magnetic domain size $w$. We emphasize that at equilibrium, analytical modeling of the magnetic thin film configuration\cite{Virot2012} predicts a decrease of $w$ upon reduction of $K_{\mathrm{u}}$. This agrees with our experimental findings: as already mentioned, a small reduction of 0.8\% of the initial $w$ value is indeed attained after 120\,ps. Assuming a homogeneous domain size reduction\cite{Pfau2012}, which is supported by the absence of any variation of $\sigma$, we can deduce a maximum domain wall velocity $v_{\mathrm{DW}} \simeq 1.3\cdot 10^4$\,m/s. Note that this is two orders of magnitude faster than current-induced DW-velocities that have been attained in Co-Tb systems so far\cite{Bang2016,Ross2018}.

\section{Conclusion}

In the present work, we have shown how tr-XRMS can be used to study the impact of optical femtosecond pulses on the magnetic properties of amorphous Co$_{88}$Tb$_{12}$ thin films exhibiting stripe domains. Our results unravel characteristic demagnetization times that are significantly smaller than what would be expected for a homogeneously magnetized thin film. This has been observed before in similar systems with large PMA and nanometer-sized domains and it was surmised, that superdiffusive spin currents might be responsible for these accelerated magnetization transients. Our present data challenge this interpretation\cite{Moisan2014}. While such currents would give rise to a sub-picosecond broadening of the domain walls, our analysis unequivocally shows that the domain wall width of the thin film remains unaffected up to $\simeq$\,4\,ps. On longer timescales, progressive energy transfer to the lattice degrees of freedom results in a successive reduction of the uniaxial anisotropy. This eventually leads to a significant increase of the domain wall width, saturating after roughly 20\,ps, the time needed by the system for full phonon equilibration through the entire thin film. 

The present results are at odds with recent work, where an ultrafast change of the domain size was observed. No such modification could be seen in our experiments. However, as pointed out in earlier studies, possible domain size shifts might sensitively depend on the pump fluence\cite{Pfau2012} which highlights the need for further systematic analysis of the experimental parameter space to provide conclusive quantitative results. 

Finally, we stress that the large discrepancy between our data and measurements on CoTb with almost identical composition but performed at different edges\cite{LopezFlores2013,Fan2019}, raises central questions concerning a possible impact of the probe energy on the ultrafast magnetic response of the system. Addressing this issue systematically could provide additional useful evidence and help to shed light on the microscopic mechanisms underlying ultrafast demagnetization. 
\newline

\section{Acknowledgments}

The authors are grateful for financial support received from the 
CNRS-MOMENTUM, the UMAMI ANR-15-CE24-0009, the CNRS-PICS and Emergence - Sorbonne Universit\'{e} programs and thank the scientific and technical teams of FERMI user facility for the support provided during the experiment. C. v. K. S would like to thank DFG for funding through TRR227 project A02. 

\section{Appendix}

\subsection{3 temperature model (3TM)}

The three temperature model has been extensively discussed in the literature and shall not be reviewed here. We follow the approach given in\cite{DallaLonga2007,Malinowski2008,Moisan2015,Vodungbo2016}. Assuming that the laser excitation
triggers an instantaneous increase of the electron temperature and further neglecting the specific heat of the spin system, one can derive the following analytical expression to describe the demagnetization process: 

\begin{widetext}
\begin{equation}
- \frac{\Delta M}{M(0)} = \Big[ (\frac{A_1}{\sqrt{\Delta t/\tau_0 +1}} - \frac{(A_2\tau_E - A_1\tau_M)}{\tau_E - \tau_M} e^{-\Delta t/\tau_M} \\- \frac{\tau_E(A_1 - A_2)}{\tau_E - \tau_M}e^{-\Delta t/\tau_E}) \theta(\Delta t)  \Big] \otimes \Gamma(\Delta t)         
\end{equation}
\end{widetext}

Here, $A_1$ represents the amplitude of the partial recovery once electrons, spins
and lattice have reached thermal equilibrium, $A_2$ described the initial magnetization
quenching, $\tau_M$ is the time constant of this quenching and $\tau_E$ is the characteristic time of the partial recovery of the magnetization, $\tau_0$ is a characteristic time describing cooling by heat diffusion. $\theta(\Delta t)$ 
is the heaviside function, used to ensure causality. Convolution with a Gaussian function $\Gamma(\Delta t)$ takes into consideration the finite pulse width.

\subsection{Translating CCD data into reciprocal space}

As described in the main part of this article, the magnetic domains act as a grating for the incoming XUV beam, giving rise to marked diffraction intensity maxima on the detector. These peaks occur at angles satisfying the equation: $n\lambda = \Lambda \sin(\theta) = \Lambda \sin\big(\mathrm{arctan}(x_n/D)\big)$, $n$ being the diffraction order, $\lambda$ the wavelength of the incoming light and $x_n$ the position of the peak on the CCD. To obtain an accurate estimation of the camera-sample distance $D$, we used static XRMS data gathered at the Co M$_{2,3}$ and Tb O$_1$ edges. This allows to calculate $D=5.9$\,cm. Data in reciprocal space is then obtained using $q = \frac{2 \pi}{\lambda} \sin\big(\mathrm{arctan}(x/D)\big)$

\subsection{1 dimensional model for analysis of the scattering pattern}

\begin{figure}
\begin{picture}(210,140)
\put(-15,0){\includegraphics[scale=1.0]{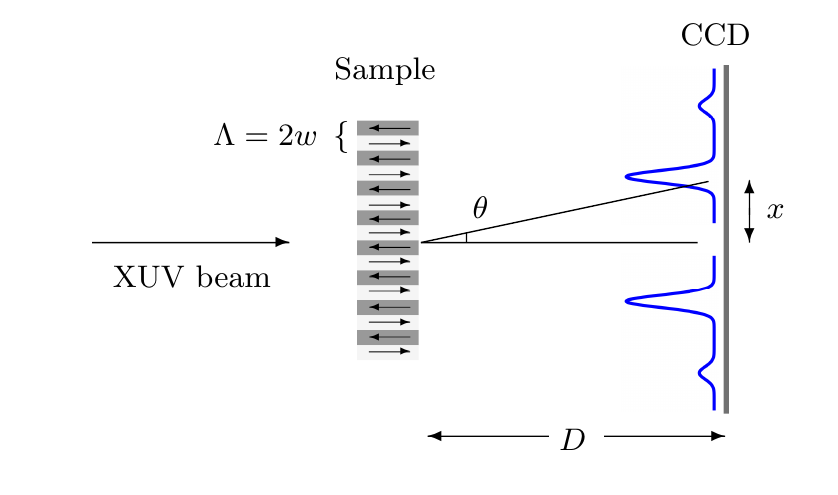}}
\end{picture}
\caption{\label{Fig:9}Simplified sketch of the scattering geometry illustrating the link between scattering peak locations on the CCD and the periodicity of the magnetic domains.}
\end{figure}

The diffraction pattern can be obtain $via$ Fourier transformation of the transmission function of the sample. The latter can be seen as a one dimensional periodic object consisting of $N$ repetitions of a basic magnetic unit cell (Fig.\,\ref{Fig:9}) with size $\Lambda$. The intensity transmitted through the sample depends on the refractive index which can be written as a function of the magnetization $M(x)$: $n(x) = 1 - \Big(\delta_0 - \delta_1M(x)\Big) + \Big(\beta_0 - \beta_1M(x)\Big)$. Using $t(x) \propto e^{i k h n(x)}$, where $h$ represents the thin film thickness, we obtain after Taylor expansion: $t(x) \propto t_0 + t_1 M(x)$, where $t_0$ and $t_1$ are constants. The transmission function is thus, up to a constant, proportional to the $z$-projection of the magnetic profile of the sample. As shown in Fig.\,10, we use a linear approximation to describe $M_z(x)$ (which, for simplicity, will be termed $\mathcal{M}(x)$ in the following). Considering the periodicity of the 1D magnetic structure, the latter can be written as the convolution of the magnetic unit cell and delta peaks with periodicity $\Lambda$

\begin{equation}
 \mathcal{M}(x) = \mathcal{M}_{\mathrm{uc}}(x) \ast \sum \delta(x - n\Lambda)
\end{equation}

In reciprocal space, the Fourier transfom of $\mathcal{M}(x)$ thus equals:

\begin{equation}
 \mathcal{M}(q) = \mathcal{M}_{\mathrm{uc}}(q) \cdot \sum_m \delta\Big(q - \frac{2 \pi m}{\Lambda}\Big)
\end{equation}

in the limit $N\rightarrow \infty$. Considering that

\begin{equation}
  \int_{-\Lambda/2}^{\Lambda/2}dx\, \mathcal{M}(x) \sin(\frac{n\pi x}{w}) = - \mathcal{M}_s \frac{8 w^2}{n^2 d \pi^2} \sin \big( \frac{n\pi d}{2 w}\big)
\end{equation}

we eventually obtain the following expression for the third and first order integrated peak intensities

\begin{equation}
 I_3/I_1 = \Big( \frac{\mathcal{M}_{\mathrm{uc,n=1}}}{\mathcal{M}_{\mathrm{uc,n=3}}} \Big)^2 = \frac{1}{81} \frac{\sin^2(3 \pi \eta/2 )}{\sin^2(\pi \eta/2)}
\end{equation}

with $\eta = d/w$, the ratio of domain wall width and domain size.

\begin{figure}
\begin{picture}(210,285)
\put(-8,0){\includegraphics[scale=1]{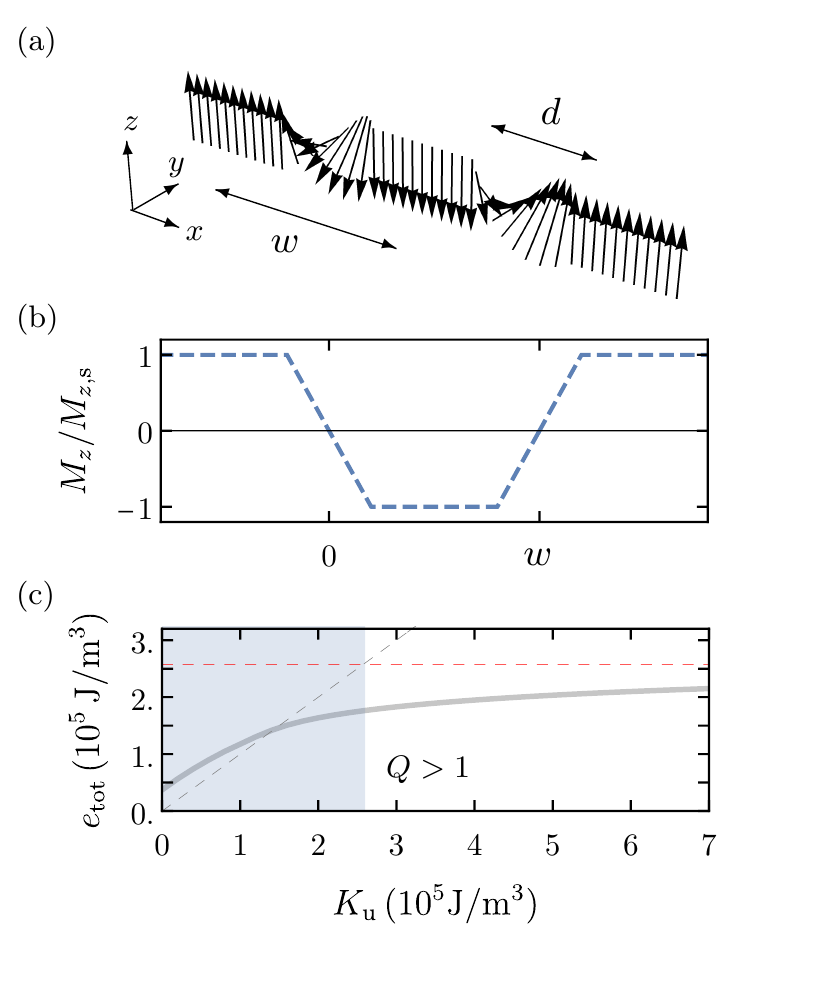}}
\end{picture}
\caption{\label{Fig:10}(a) One dimensional spin chain with domains of size $w$ and Bloch walls (width $d$) and (b) linear approximation used to describe the normalized $z$-component of the magnetization $M_{\mathrm{s}}$. (c) Total magnetic energy density of the striped thin film as a function of anisotropy (gray solid line). The magnetostatic contribution $\frac{1}{2} \mu_0 M^2_{\mathrm{s}}$ is shown as a red dashed line. For $Q = \frac{2 K}{\mu_0 M^2_{\mathrm{s}}} > 1$, the equilibrium magnetization will point in out of plane direction, irrespective of the thin film thickness.}
\end{figure}

\subsection{Analytical description of magnetic stripe structures}

The creation of magnetic stripe structures has initially been treated by Kittel\cite{Kittel1949}. However, his model is limited to the description of systems, where the domain wall width can be neglected, when compared to the domain size ($d \ll w$) and where the domain size is much smaller than the film thickness. This does not hold in our CoTb thin films. In the present work, we thus sticked to the sine wave model (SWM) proposed by Virot \textit{et al.}, which permits to circumvent these problems\cite{Virot2012}. The total energy of the thin film is written as a sum of magnetostatic, anisotropy and exchange contributions:

\begin{equation}
 e_{\mathrm{tot}} = e_{\mathrm{ms}} + e_{K} + e_{A}
\end{equation}

The magnetostatic energy per unit volume is equal to

\begin{equation}
 e_{\mathrm{ms}} = \frac{\mu_0 w}{\pi h} \sum_{k\,\mathrm{odd}}^{\infty} \frac{|C_k|^2}{k}\Big[1 - e^{-\pi k h/w} \Big]
\end{equation}

with Fourier coefficients

\begin{equation}
C_k = \frac{2 M_{\mathrm{s}}}{k \pi [1 - k^2 (d/w)^2]} \cos\Big(\frac{k \pi d}{2 w}\Big)
\end{equation}

The exchange contribution equals

\begin{equation}
 e_{A} = \frac{\pi^2 A}{d w} 
\end{equation}

while the anisotropy energy density is written as

\begin{equation}
 e_{K} = \sum_i \frac{1}{2w} \int_0^{2w} K_{\mathrm{u},i} \Big[\cos\big(\vartheta(x)\big)\Big]^{2i} dx 
\end{equation}

where $\vartheta(x)$ describes the angle between the local magnetization and the $y$-axis, changing linearly along the domain wall width (see Fig.\,10). In their paper, Virot \textit{et al.} only consider $i=1$. 
%which gives rise to $e_{K_u} = \frac{K_ud}{2w}$. Here, 
%the magnetometry measurements clearly hint at higher order contributions, which would at least require the calculation of 
we additionally take into consideration the next order term ($i=2$).
%to adequately describe the observed non-linear IP $M(H)$ relationship when approaching saturation. 
However, considering $K_{u,2}$ to be smaller than $K_{u,1}$, the original first order result of Virot \textit{et al.} $e_{K} = \frac{K_u d}{2w}$ can still be used with $K_{u} = K_{u,1} + K_{u,2}$. In the present case, it represents a reasonable approximation to the anisotropy energy density (less than 6\% relative error for $K_{u,2} < K_{u,1}/3$). Using the aforementioned energy contributions, $e_{\mathrm{tot}}$ is obtained by numerical minimization of eq.\,(3) with respect to $w$ and $d$. This eventually allows to obtain the equilibrium values $w(K)$, $d(K)$ shown in Fig.\,\ref{Fig:7}.  Note that the calculation of $e_{K}$ requires knowledge of $A$, the exchange stiffness of the system. We calculated this quantity via\cite{Hasegawa1974}

\begin{equation}
 A = \sum_{i,j} 2|J_{ij}| P_{ij} S_iS_j  /a_{ij}
\end{equation}

Here, the two sums are taken over the atomic species, Co and Tb, respectively. The $J_{ij}$ represent the exchange integrals ($J_{\mathrm{CoCo}} = 2.1\cdot 10^{-21}$\,J, $J_{\mathrm{CoTb}} = -2.4\cdot 10^{-22}$\,J and $ J_{\mathrm{TbTb}} = 0.2\cdot 10^{-22}$\,J)\cite{Zou2003}. $P_{ij}$ corresponds to the probability of having a given atomic pair $ij$ and can thus be linked to the concentration (considering that the alloy is randomly mixed), $S_i$ is the average spin value of species $i$ ($S_{\mathrm{Co}} = 0.73$ and $S_{\mathrm{Tb}} = 5.05$)\cite{Zou2003} while $a_{ij}$ denotes the average interatomic distances between atoms of type $i$ and $j$ ($a_{\mathrm{CoCo}} = 2.5$\,\AA, $a_{\mathrm{CoTb}} = 3.0$\,\AA\,\,and $a_{\mathrm{TbTb}} = 3.5$\,\AA)\cite{}. With these data, we obtain $A \simeq 1.0\cdot 10^{-11}$\,J/m.

\bibliography{ala}% Produces the bibliography via BibTeX.

%merlin.mbs apsrev4-1.bst 2010-07-25 4.21a (PWD, AO, DPC) hacked
%Control: key (0)
%Control: author (8) initials jnrlst
%Control: editor formatted (1) identically to author
%Control: production of article title (-1) disabled
%Control: page (0) single
%Control: year (1) truncated
%Control: production of eprint (0) enabled
\begin{thebibliography}{59}%
\makeatletter
\providecommand \@ifxundefined [1]{%
 \@ifx{#1\undefined}
}%
\providecommand \@ifnum [1]{%
 \ifnum #1\expandafter \@firstoftwo
 \else \expandafter \@secondoftwo
 \fi
}%
\providecommand \@ifx [1]{%
 \ifx #1\expandafter \@firstoftwo
 \else \expandafter \@secondoftwo
 \fi
}%
\providecommand \natexlab [1]{#1}%
\providecommand \enquote  [1]{``#1''}%
\providecommand \bibnamefont  [1]{#1}%
\providecommand \bibfnamefont [1]{#1}%
\providecommand \citenamefont [1]{#1}%
\providecommand \href@noop [0]{\@secondoftwo}%
\providecommand \href [0]{\begingroup \@sanitize@url \@href}%
\providecommand \@href[1]{\@@startlink{#1}\@@href}%
\providecommand \@@href[1]{\endgroup#1\@@endlink}%
\providecommand \@sanitize@url [0]{\catcode `\\12\catcode `\$12\catcode
  `\&12\catcode `\#12\catcode `\^12\catcode `\_12\catcode `\%12\relax}%
\providecommand \@@startlink[1]{}%
\providecommand \@@endlink[0]{}%
\providecommand \url  [0]{\begingroup\@sanitize@url \@url }%
\providecommand \@url [1]{\endgroup\@href {#1}{\urlprefix }}%
\providecommand \urlprefix  [0]{URL }%
\providecommand \Eprint [0]{\href }%
\providecommand \doibase [0]{http://dx.doi.org/}%
\providecommand \selectlanguage [0]{\@gobble}%
\providecommand \bibinfo  [0]{\@secondoftwo}%
\providecommand \bibfield  [0]{\@secondoftwo}%
\providecommand \translation [1]{[#1]}%
\providecommand \BibitemOpen [0]{}%
\providecommand \bibitemStop [0]{}%
\providecommand \bibitemNoStop [0]{.\EOS\space}%
\providecommand \EOS [0]{\spacefactor3000\relax}%
\providecommand \BibitemShut  [1]{\csname bibitem#1\endcsname}%
\let\auto@bib@innerbib\@empty
%</preamble>
\bibitem [{\citenamefont {Beaurepaire}\ \emph {et~al.}(1996)\citenamefont
  {Beaurepaire}, \citenamefont {Merle}, \citenamefont {Daunois},\ and\
  \citenamefont {Bigot}}]{Beaurepaire1996}%
  \BibitemOpen
  \bibfield  {author} {\bibinfo {author} {\bibfnamefont {E.}~\bibnamefont
  {Beaurepaire}}, \bibinfo {author} {\bibfnamefont {J.-C.}\ \bibnamefont
  {Merle}}, \bibinfo {author} {\bibfnamefont {A.}~\bibnamefont {Daunois}}, \
  and\ \bibinfo {author} {\bibfnamefont {J.-Y.}\ \bibnamefont {Bigot}},\ }\href
  {\doibase 10.1103/PhysRevLett.76.4250} {\bibfield  {journal} {\bibinfo
  {journal} {Phys. Rev. Lett.}\ }\textbf {\bibinfo {volume} {76}},\ \bibinfo
  {pages} {4250} (\bibinfo {year} {1996})}\BibitemShut {NoStop}%
\bibitem [{\citenamefont {Carpene}\ \emph {et~al.}(2008)\citenamefont
  {Carpene}, \citenamefont {Mancini}, \citenamefont {Dallera}, \citenamefont
  {Brenna}, \citenamefont {Puppin},\ and\ \citenamefont
  {De~Silvestri}}]{Carpene2008}%
  \BibitemOpen
  \bibfield  {author} {\bibinfo {author} {\bibfnamefont {E.}~\bibnamefont
  {Carpene}}, \bibinfo {author} {\bibfnamefont {E.}~\bibnamefont {Mancini}},
  \bibinfo {author} {\bibfnamefont {C.}~\bibnamefont {Dallera}}, \bibinfo
  {author} {\bibfnamefont {M.}~\bibnamefont {Brenna}}, \bibinfo {author}
  {\bibfnamefont {E.}~\bibnamefont {Puppin}}, \ and\ \bibinfo {author}
  {\bibfnamefont {S.}~\bibnamefont {De~Silvestri}},\ }\href {\doibase
  10.1103/PhysRevB.78.174422} {\bibfield  {journal} {\bibinfo  {journal} {Phys.
  Rev. B}\ }\textbf {\bibinfo {volume} {78}},\ \bibinfo {pages} {174422}
  (\bibinfo {year} {2008})}\BibitemShut {NoStop}%
\bibitem [{\citenamefont {Krau\ss{}}\ \emph {et~al.}(2009)\citenamefont
  {Krau\ss{}}, \citenamefont {Roth}, \citenamefont {Alebrand}, \citenamefont
  {Steil}, \citenamefont {Cinchetti}, \citenamefont {Aeschlimann},\ and\
  \citenamefont {Schneider}}]{Krauss2009}%
  \BibitemOpen
  \bibfield  {author} {\bibinfo {author} {\bibfnamefont {M.}~\bibnamefont
  {Krau\ss{}}}, \bibinfo {author} {\bibfnamefont {T.}~\bibnamefont {Roth}},
  \bibinfo {author} {\bibfnamefont {S.}~\bibnamefont {Alebrand}}, \bibinfo
  {author} {\bibfnamefont {D.}~\bibnamefont {Steil}}, \bibinfo {author}
  {\bibfnamefont {M.}~\bibnamefont {Cinchetti}}, \bibinfo {author}
  {\bibfnamefont {M.}~\bibnamefont {Aeschlimann}}, \ and\ \bibinfo {author}
  {\bibfnamefont {H.~C.}\ \bibnamefont {Schneider}},\ }\href {\doibase
  10.1103/PhysRevB.80.180407} {\bibfield  {journal} {\bibinfo  {journal} {Phys.
  Rev. B}\ }\textbf {\bibinfo {volume} {80}},\ \bibinfo {pages} {180407}
  (\bibinfo {year} {2009})}\BibitemShut {NoStop}%
\bibitem [{\citenamefont {Bigot}\ \emph {et~al.}(2009)\citenamefont {Bigot},
  \citenamefont {Vomir},\ and\ \citenamefont {Beaurepaire}}]{Bigot2009}%
  \BibitemOpen
  \bibfield  {author} {\bibinfo {author} {\bibfnamefont {J.-Y.}\ \bibnamefont
  {Bigot}}, \bibinfo {author} {\bibfnamefont {M.}~\bibnamefont {Vomir}}, \ and\
  \bibinfo {author} {\bibfnamefont {E.}~\bibnamefont {Beaurepaire}},\ }\href
  {\doibase 10.1038/nphys1285} {\bibfield  {journal} {\bibinfo  {journal}
  {Nature Physics}\ }\textbf {\bibinfo {volume} {5}},\ \bibinfo {pages} {515}
  (\bibinfo {year} {2009})}\BibitemShut {NoStop}%
\bibitem [{\citenamefont {Koopmans}\ \emph {et~al.}(2010)\citenamefont
  {Koopmans}, \citenamefont {Malinowski}, \citenamefont {Dalla~Longa},
  \citenamefont {Steiauf}, \citenamefont {F{\"a}hnle}, \citenamefont {Roth},
  \citenamefont {Cinchetti},\ and\ \citenamefont {Aeschlimann}}]{Koopmans2010}%
  \BibitemOpen
  \bibfield  {author} {\bibinfo {author} {\bibfnamefont {B.}~\bibnamefont
  {Koopmans}}, \bibinfo {author} {\bibfnamefont {G.}~\bibnamefont
  {Malinowski}}, \bibinfo {author} {\bibfnamefont {F.}~\bibnamefont
  {Dalla~Longa}}, \bibinfo {author} {\bibfnamefont {D.}~\bibnamefont
  {Steiauf}}, \bibinfo {author} {\bibfnamefont {M.}~\bibnamefont {F{\"a}hnle}},
  \bibinfo {author} {\bibfnamefont {T.}~\bibnamefont {Roth}}, \bibinfo {author}
  {\bibfnamefont {M.}~\bibnamefont {Cinchetti}}, \ and\ \bibinfo {author}
  {\bibfnamefont {M.}~\bibnamefont {Aeschlimann}},\ }\href {\doibase
  10.1038/nmat2593} {\bibfield  {journal} {\bibinfo  {journal} {Nature
  Materials}\ }\textbf {\bibinfo {volume} {9}},\ \bibinfo {pages} {259}
  (\bibinfo {year} {2010})}\BibitemShut {NoStop}%
\bibitem [{\citenamefont {Battiato}\ \emph {et~al.}(2010)\citenamefont
  {Battiato}, \citenamefont {Carva},\ and\ \citenamefont
  {Oppeneer}}]{Battiato2010}%
  \BibitemOpen
  \bibfield  {author} {\bibinfo {author} {\bibfnamefont {M.}~\bibnamefont
  {Battiato}}, \bibinfo {author} {\bibfnamefont {K.}~\bibnamefont {Carva}}, \
  and\ \bibinfo {author} {\bibfnamefont {P.~M.}\ \bibnamefont {Oppeneer}},\
  }\href {\doibase 10.1103/PhysRevLett.105.027203} {\bibfield  {journal}
  {\bibinfo  {journal} {Phys. Rev. Lett.}\ }\textbf {\bibinfo {volume} {105}},\
  \bibinfo {pages} {027203} (\bibinfo {year} {2010})}\BibitemShut {NoStop}%
\bibitem [{\citenamefont {Illg}\ \emph {et~al.}(2013)\citenamefont {Illg},
  \citenamefont {Haag},\ and\ \citenamefont {F\"ahnle}}]{Illg2013}%
  \BibitemOpen
  \bibfield  {author} {\bibinfo {author} {\bibfnamefont {C.}~\bibnamefont
  {Illg}}, \bibinfo {author} {\bibfnamefont {M.}~\bibnamefont {Haag}}, \ and\
  \bibinfo {author} {\bibfnamefont {M.}~\bibnamefont {F\"ahnle}},\ }\href
  {\doibase 10.1103/PhysRevB.88.214404} {\bibfield  {journal} {\bibinfo
  {journal} {Phys. Rev. B}\ }\textbf {\bibinfo {volume} {88}},\ \bibinfo
  {pages} {214404} (\bibinfo {year} {2013})}\BibitemShut {NoStop}%
\bibitem [{\citenamefont {Zhang}\ \emph {et~al.}(2018)\citenamefont {Zhang},
  \citenamefont {Bai}, \citenamefont {Jenkins},\ and\ \citenamefont
  {George}}]{Zhang2018}%
  \BibitemOpen
  \bibfield  {author} {\bibinfo {author} {\bibfnamefont {G.~P.}\ \bibnamefont
  {Zhang}}, \bibinfo {author} {\bibfnamefont {Y.~H.}\ \bibnamefont {Bai}},
  \bibinfo {author} {\bibfnamefont {T.}~\bibnamefont {Jenkins}}, \ and\
  \bibinfo {author} {\bibfnamefont {T.~F.}\ \bibnamefont {George}},\ }\href
  {\doibase 10.1088/1361-648x/aae5a9} {\bibfield  {journal} {\bibinfo
  {journal} {Journal of Physics: Condensed Matter}\ }\textbf {\bibinfo {volume}
  {30}},\ \bibinfo {pages} {465801} (\bibinfo {year} {2018})}\BibitemShut
  {NoStop}%
\bibitem [{\citenamefont {Dewhurst}\ \emph {et~al.}(2018)\citenamefont
  {Dewhurst}, \citenamefont {Elliott}, \citenamefont {Shallcross},
  \citenamefont {Gross},\ and\ \citenamefont {Sharma}}]{Dewhurst2018}%
  \BibitemOpen
  \bibfield  {author} {\bibinfo {author} {\bibfnamefont {J.~K.}\ \bibnamefont
  {Dewhurst}}, \bibinfo {author} {\bibfnamefont {P.}~\bibnamefont {Elliott}},
  \bibinfo {author} {\bibfnamefont {S.}~\bibnamefont {Shallcross}}, \bibinfo
  {author} {\bibfnamefont {E.~K.~U.}\ \bibnamefont {Gross}}, \ and\ \bibinfo
  {author} {\bibfnamefont {S.}~\bibnamefont {Sharma}},\ }\href {\doibase
  10.1021/acs.nanolett.7b05118} {\bibfield  {journal} {\bibinfo  {journal}
  {Nano Letters}\ }\textbf {\bibinfo {volume} {18}},\ \bibinfo {pages} {1842}
  (\bibinfo {year} {2018})}\BibitemShut {NoStop}%
\bibitem [{\citenamefont {Dornes}\ \emph {et~al.}(2019)\citenamefont {Dornes},
  \citenamefont {Acremann}, \citenamefont {Savoini}, \citenamefont {Kubli},
  \citenamefont {Neugebauer}, \citenamefont {Abreu}, \citenamefont {Huber},
  \citenamefont {Lantz}, \citenamefont {Vaz}, \citenamefont {Lemke},
  \citenamefont {Bothschafter}, \citenamefont {Porer}, \citenamefont
  {Esposito}, \citenamefont {Rettig}, \citenamefont {Buzzi}, \citenamefont
  {Alberca}, \citenamefont {Windsor}, \citenamefont {Beaud}, \citenamefont
  {Staub}, \citenamefont {Zhu}, \citenamefont {Song}, \citenamefont {Glownia},\
  and\ \citenamefont {Johnson}}]{Dornes2019}%
  \BibitemOpen
  \bibfield  {author} {\bibinfo {author} {\bibfnamefont {C.}~\bibnamefont
  {Dornes}}, \bibinfo {author} {\bibfnamefont {Y.}~\bibnamefont {Acremann}},
  \bibinfo {author} {\bibfnamefont {M.}~\bibnamefont {Savoini}}, \bibinfo
  {author} {\bibfnamefont {M.}~\bibnamefont {Kubli}}, \bibinfo {author}
  {\bibfnamefont {M.~J.}\ \bibnamefont {Neugebauer}}, \bibinfo {author}
  {\bibfnamefont {E.}~\bibnamefont {Abreu}}, \bibinfo {author} {\bibfnamefont
  {L.}~\bibnamefont {Huber}}, \bibinfo {author} {\bibfnamefont
  {G.}~\bibnamefont {Lantz}}, \bibinfo {author} {\bibfnamefont {C.~A.~F.}\
  \bibnamefont {Vaz}}, \bibinfo {author} {\bibfnamefont {H.}~\bibnamefont
  {Lemke}}, \bibinfo {author} {\bibfnamefont {E.~M.}\ \bibnamefont
  {Bothschafter}}, \bibinfo {author} {\bibfnamefont {M.}~\bibnamefont {Porer}},
  \bibinfo {author} {\bibfnamefont {V.}~\bibnamefont {Esposito}}, \bibinfo
  {author} {\bibfnamefont {L.}~\bibnamefont {Rettig}}, \bibinfo {author}
  {\bibfnamefont {M.}~\bibnamefont {Buzzi}}, \bibinfo {author} {\bibfnamefont
  {A.}~\bibnamefont {Alberca}}, \bibinfo {author} {\bibfnamefont {Y.~W.}\
  \bibnamefont {Windsor}}, \bibinfo {author} {\bibfnamefont {P.}~\bibnamefont
  {Beaud}}, \bibinfo {author} {\bibfnamefont {U.}~\bibnamefont {Staub}},
  \bibinfo {author} {\bibfnamefont {D.}~\bibnamefont {Zhu}}, \bibinfo {author}
  {\bibfnamefont {S.}~\bibnamefont {Song}}, \bibinfo {author} {\bibfnamefont
  {J.~M.}\ \bibnamefont {Glownia}}, \ and\ \bibinfo {author} {\bibfnamefont
  {S.~L.}\ \bibnamefont {Johnson}},\ }\href {\doibase
  10.1038/s41586-018-0822-7} {\bibfield  {journal} {\bibinfo  {journal}
  {Nature}\ }\textbf {\bibinfo {volume} {565}},\ \bibinfo {pages} {209}
  (\bibinfo {year} {2019})}\BibitemShut {NoStop}%
\bibitem [{\citenamefont {Battiato}\ \emph {et~al.}(2012)\citenamefont
  {Battiato}, \citenamefont {Carva},\ and\ \citenamefont
  {Oppeneer}}]{Battiato2012}%
  \BibitemOpen
  \bibfield  {author} {\bibinfo {author} {\bibfnamefont {M.}~\bibnamefont
  {Battiato}}, \bibinfo {author} {\bibfnamefont {K.}~\bibnamefont {Carva}}, \
  and\ \bibinfo {author} {\bibfnamefont {P.~M.}\ \bibnamefont {Oppeneer}},\
  }\href {\doibase 10.1103/PhysRevB.86.024404} {\bibfield  {journal} {\bibinfo
  {journal} {Phys. Rev. B}\ }\textbf {\bibinfo {volume} {86}},\ \bibinfo
  {pages} {024404} (\bibinfo {year} {2012})}\BibitemShut {NoStop}%
\bibitem [{\citenamefont {Malinowski}\ \emph {et~al.}(2008)\citenamefont
  {Malinowski}, \citenamefont {Dalla~Longa}, \citenamefont {Rietjens},
  \citenamefont {Paluskar}, \citenamefont {Huijink}, \citenamefont {Swagten},\
  and\ \citenamefont {Koopmans}}]{Malinowski2008}%
  \BibitemOpen
  \bibfield  {author} {\bibinfo {author} {\bibfnamefont {G.}~\bibnamefont
  {Malinowski}}, \bibinfo {author} {\bibfnamefont {F.}~\bibnamefont
  {Dalla~Longa}}, \bibinfo {author} {\bibfnamefont {J.~H.~H.}\ \bibnamefont
  {Rietjens}}, \bibinfo {author} {\bibfnamefont {P.~V.}\ \bibnamefont
  {Paluskar}}, \bibinfo {author} {\bibfnamefont {R.}~\bibnamefont {Huijink}},
  \bibinfo {author} {\bibfnamefont {H.~J.~M.}\ \bibnamefont {Swagten}}, \ and\
  \bibinfo {author} {\bibfnamefont {B.}~\bibnamefont {Koopmans}},\ }\href
  {https://doi.org/10.1038/nphys1092} {\bibfield  {journal} {\bibinfo
  {journal} {Nature Physics}\ }\textbf {\bibinfo {volume} {4}},\ \bibinfo
  {pages} {855 EP } (\bibinfo {year} {2008})}\BibitemShut {NoStop}%
\bibitem [{\citenamefont {Vodungbo}\ \emph {et~al.}(2012)\citenamefont
  {Vodungbo}, \citenamefont {Gautier}, \citenamefont {Lambert}, \citenamefont
  {Sardinha}, \citenamefont {Lozano}, \citenamefont {Sebban}, \citenamefont
  {Ducousso}, \citenamefont {Boutu}, \citenamefont {Li}, \citenamefont {Tudu},
  \citenamefont {Tortarolo}, \citenamefont {Hawaldar}, \citenamefont
  {Delaunay}, \citenamefont {L{\'o}pez-Flores}, \citenamefont {Arabski},
  \citenamefont {Boeglin}, \citenamefont {Merdji}, \citenamefont {Zeitoun},\
  and\ \citenamefont {L{\"u}ning}}]{Vodungbo2012}%
  \BibitemOpen
  \bibfield  {author} {\bibinfo {author} {\bibfnamefont {B.}~\bibnamefont
  {Vodungbo}}, \bibinfo {author} {\bibfnamefont {J.}~\bibnamefont {Gautier}},
  \bibinfo {author} {\bibfnamefont {G.}~\bibnamefont {Lambert}}, \bibinfo
  {author} {\bibfnamefont {A.~B.}\ \bibnamefont {Sardinha}}, \bibinfo {author}
  {\bibfnamefont {M.}~\bibnamefont {Lozano}}, \bibinfo {author} {\bibfnamefont
  {S.}~\bibnamefont {Sebban}}, \bibinfo {author} {\bibfnamefont
  {M.}~\bibnamefont {Ducousso}}, \bibinfo {author} {\bibfnamefont
  {W.}~\bibnamefont {Boutu}}, \bibinfo {author} {\bibfnamefont
  {K.}~\bibnamefont {Li}}, \bibinfo {author} {\bibfnamefont {B.}~\bibnamefont
  {Tudu}}, \bibinfo {author} {\bibfnamefont {M.}~\bibnamefont {Tortarolo}},
  \bibinfo {author} {\bibfnamefont {R.}~\bibnamefont {Hawaldar}}, \bibinfo
  {author} {\bibfnamefont {R.}~\bibnamefont {Delaunay}}, \bibinfo {author}
  {\bibfnamefont {V.}~\bibnamefont {L{\'o}pez-Flores}}, \bibinfo {author}
  {\bibfnamefont {J.}~\bibnamefont {Arabski}}, \bibinfo {author} {\bibfnamefont
  {C.}~\bibnamefont {Boeglin}}, \bibinfo {author} {\bibfnamefont
  {H.}~\bibnamefont {Merdji}}, \bibinfo {author} {\bibfnamefont
  {P.}~\bibnamefont {Zeitoun}}, \ and\ \bibinfo {author} {\bibfnamefont
  {J.}~\bibnamefont {L{\"u}ning}},\ }\href {https://doi.org/10.1038/ncomms2007}
  {\bibfield  {journal} {\bibinfo  {journal} {Nature Communications}\ }\textbf
  {\bibinfo {volume} {3}},\ \bibinfo {pages} {999 EP } (\bibinfo {year}
  {2012})},\ \bibinfo {note} {article}\BibitemShut {NoStop}%
\bibitem [{\citenamefont {Rudolf}\ \emph {et~al.}(2012)\citenamefont {Rudolf},
  \citenamefont {La-O-Vorakiat}, \citenamefont {Battiato}, \citenamefont
  {Adam}, \citenamefont {Shaw}, \citenamefont {Turgut}, \citenamefont
  {Maldonado}, \citenamefont {Mathias}, \citenamefont {Grychtol}, \citenamefont
  {Nembach}, \citenamefont {Silva}, \citenamefont {Aeschlimann}, \citenamefont
  {Kapteyn}, \citenamefont {Murnane}, \citenamefont {Schneider},\ and\
  \citenamefont {Oppeneer}}]{Rudolf2012}%
  \BibitemOpen
  \bibfield  {author} {\bibinfo {author} {\bibfnamefont {D.}~\bibnamefont
  {Rudolf}}, \bibinfo {author} {\bibfnamefont {C.}~\bibnamefont
  {La-O-Vorakiat}}, \bibinfo {author} {\bibfnamefont {M.}~\bibnamefont
  {Battiato}}, \bibinfo {author} {\bibfnamefont {R.}~\bibnamefont {Adam}},
  \bibinfo {author} {\bibfnamefont {J.~M.}\ \bibnamefont {Shaw}}, \bibinfo
  {author} {\bibfnamefont {E.}~\bibnamefont {Turgut}}, \bibinfo {author}
  {\bibfnamefont {P.}~\bibnamefont {Maldonado}}, \bibinfo {author}
  {\bibfnamefont {S.}~\bibnamefont {Mathias}}, \bibinfo {author} {\bibfnamefont
  {P.}~\bibnamefont {Grychtol}}, \bibinfo {author} {\bibfnamefont {H.~T.}\
  \bibnamefont {Nembach}}, \bibinfo {author} {\bibfnamefont {T.~J.}\
  \bibnamefont {Silva}}, \bibinfo {author} {\bibfnamefont {M.}~\bibnamefont
  {Aeschlimann}}, \bibinfo {author} {\bibfnamefont {H.~C.}\ \bibnamefont
  {Kapteyn}}, \bibinfo {author} {\bibfnamefont {M.~M.}\ \bibnamefont
  {Murnane}}, \bibinfo {author} {\bibfnamefont {C.~M.}\ \bibnamefont
  {Schneider}}, \ and\ \bibinfo {author} {\bibfnamefont {P.~M.}\ \bibnamefont
  {Oppeneer}},\ }\href {\doibase 10.1038/ncomms2029} {\bibfield  {journal}
  {\bibinfo  {journal} {Nature Communications}\ }\textbf {\bibinfo {volume}
  {3}},\ \bibinfo {pages} {1037} (\bibinfo {year} {2012})}\BibitemShut
  {NoStop}%
\bibitem [{\citenamefont {Pfau}\ \emph {et~al.}(2012)\citenamefont {Pfau},
  \citenamefont {Schaffert}, \citenamefont {M{\"u}ller}, \citenamefont {Gutt},
  \citenamefont {Al-Shemmary}, \citenamefont {B{\"u}ttner}, \citenamefont
  {Delaunay}, \citenamefont {D{\"u}sterer}, \citenamefont {Flewett},
  \citenamefont {Fr{\"o}mter}, \citenamefont {Geilhufe}, \citenamefont
  {Guehrs}, \citenamefont {G{\"u}nther}, \citenamefont {Hawaldar},
  \citenamefont {Hille}, \citenamefont {Jaouen}, \citenamefont {Kobs},
  \citenamefont {Li}, \citenamefont {Mohanty}, \citenamefont {Redlin},
  \citenamefont {Schlotter}, \citenamefont {Stickler}, \citenamefont {Treusch},
  \citenamefont {Vodungbo}, \citenamefont {Kl{\"a}ui}, \citenamefont {Oepen},
  \citenamefont {L{\"u}ning}, \citenamefont {Gr{\"u}bel},\ and\ \citenamefont
  {Eisebitt}}]{Pfau2012}%
  \BibitemOpen
  \bibfield  {author} {\bibinfo {author} {\bibfnamefont {B.}~\bibnamefont
  {Pfau}}, \bibinfo {author} {\bibfnamefont {S.}~\bibnamefont {Schaffert}},
  \bibinfo {author} {\bibfnamefont {L.}~\bibnamefont {M{\"u}ller}}, \bibinfo
  {author} {\bibfnamefont {C.}~\bibnamefont {Gutt}}, \bibinfo {author}
  {\bibfnamefont {A.}~\bibnamefont {Al-Shemmary}}, \bibinfo {author}
  {\bibfnamefont {F.}~\bibnamefont {B{\"u}ttner}}, \bibinfo {author}
  {\bibfnamefont {R.}~\bibnamefont {Delaunay}}, \bibinfo {author}
  {\bibfnamefont {S.}~\bibnamefont {D{\"u}sterer}}, \bibinfo {author}
  {\bibfnamefont {S.}~\bibnamefont {Flewett}}, \bibinfo {author} {\bibfnamefont
  {R.}~\bibnamefont {Fr{\"o}mter}}, \bibinfo {author} {\bibfnamefont
  {J.}~\bibnamefont {Geilhufe}}, \bibinfo {author} {\bibfnamefont
  {E.}~\bibnamefont {Guehrs}}, \bibinfo {author} {\bibfnamefont {C.~M.}\
  \bibnamefont {G{\"u}nther}}, \bibinfo {author} {\bibfnamefont
  {R.}~\bibnamefont {Hawaldar}}, \bibinfo {author} {\bibfnamefont
  {M.}~\bibnamefont {Hille}}, \bibinfo {author} {\bibfnamefont
  {N.}~\bibnamefont {Jaouen}}, \bibinfo {author} {\bibfnamefont
  {A.}~\bibnamefont {Kobs}}, \bibinfo {author} {\bibfnamefont {K.}~\bibnamefont
  {Li}}, \bibinfo {author} {\bibfnamefont {J.}~\bibnamefont {Mohanty}},
  \bibinfo {author} {\bibfnamefont {H.}~\bibnamefont {Redlin}}, \bibinfo
  {author} {\bibfnamefont {W.~F.}\ \bibnamefont {Schlotter}}, \bibinfo {author}
  {\bibfnamefont {D.}~\bibnamefont {Stickler}}, \bibinfo {author}
  {\bibfnamefont {R.}~\bibnamefont {Treusch}}, \bibinfo {author} {\bibfnamefont
  {B.}~\bibnamefont {Vodungbo}}, \bibinfo {author} {\bibfnamefont
  {M.}~\bibnamefont {Kl{\"a}ui}}, \bibinfo {author} {\bibfnamefont {H.~P.}\
  \bibnamefont {Oepen}}, \bibinfo {author} {\bibfnamefont {J.}~\bibnamefont
  {L{\"u}ning}}, \bibinfo {author} {\bibfnamefont {G.}~\bibnamefont
  {Gr{\"u}bel}}, \ and\ \bibinfo {author} {\bibfnamefont {S.}~\bibnamefont
  {Eisebitt}},\ }\href {\doibase 10.1038/ncomms2108} {\bibfield  {journal}
  {\bibinfo  {journal} {Nature Communications}\ }\textbf {\bibinfo {volume}
  {3}},\ \bibinfo {pages} {1100} (\bibinfo {year} {2012})}\BibitemShut
  {NoStop}%
\bibitem [{\citenamefont {Graves}\ \emph {et~al.}(2013)\citenamefont {Graves},
  \citenamefont {Reid}, \citenamefont {Wang}, \citenamefont {Wu}, \citenamefont
  {de~Jong}, \citenamefont {Vahaplar}, \citenamefont {Radu}, \citenamefont
  {Bernstein}, \citenamefont {Messerschmidt}, \citenamefont {M{\"u}ller},
  \citenamefont {Coffee}, \citenamefont {Bionta}, \citenamefont {Epp},
  \citenamefont {Hartmann}, \citenamefont {Kimmel}, \citenamefont {Hauser},
  \citenamefont {Hartmann}, \citenamefont {Holl}, \citenamefont {Gorke},
  \citenamefont {Mentink}, \citenamefont {Tsukamoto}, \citenamefont {Fognini},
  \citenamefont {Turner}, \citenamefont {Schlotter}, \citenamefont {Rolles},
  \citenamefont {Soltau}, \citenamefont {Str{\"u}der}, \citenamefont
  {Acremann}, \citenamefont {Kimel}, \citenamefont {Kirilyuk}, \citenamefont
  {Rasing}, \citenamefont {St{\"o}hr}, \citenamefont {Scherz},\ and\
  \citenamefont {D{\"u}rr}}]{Graves2013}%
  \BibitemOpen
  \bibfield  {author} {\bibinfo {author} {\bibfnamefont {C.~E.}\ \bibnamefont
  {Graves}}, \bibinfo {author} {\bibfnamefont {A.~H.}\ \bibnamefont {Reid}},
  \bibinfo {author} {\bibfnamefont {T.}~\bibnamefont {Wang}}, \bibinfo {author}
  {\bibfnamefont {B.}~\bibnamefont {Wu}}, \bibinfo {author} {\bibfnamefont
  {S.}~\bibnamefont {de~Jong}}, \bibinfo {author} {\bibfnamefont
  {K.}~\bibnamefont {Vahaplar}}, \bibinfo {author} {\bibfnamefont
  {I.}~\bibnamefont {Radu}}, \bibinfo {author} {\bibfnamefont {D.~P.}\
  \bibnamefont {Bernstein}}, \bibinfo {author} {\bibfnamefont {M.}~\bibnamefont
  {Messerschmidt}}, \bibinfo {author} {\bibfnamefont {L.}~\bibnamefont
  {M{\"u}ller}}, \bibinfo {author} {\bibfnamefont {R.}~\bibnamefont {Coffee}},
  \bibinfo {author} {\bibfnamefont {M.}~\bibnamefont {Bionta}}, \bibinfo
  {author} {\bibfnamefont {S.~W.}\ \bibnamefont {Epp}}, \bibinfo {author}
  {\bibfnamefont {R.}~\bibnamefont {Hartmann}}, \bibinfo {author}
  {\bibfnamefont {N.}~\bibnamefont {Kimmel}}, \bibinfo {author} {\bibfnamefont
  {G.}~\bibnamefont {Hauser}}, \bibinfo {author} {\bibfnamefont
  {A.}~\bibnamefont {Hartmann}}, \bibinfo {author} {\bibfnamefont
  {P.}~\bibnamefont {Holl}}, \bibinfo {author} {\bibfnamefont {H.}~\bibnamefont
  {Gorke}}, \bibinfo {author} {\bibfnamefont {J.~H.}\ \bibnamefont {Mentink}},
  \bibinfo {author} {\bibfnamefont {A.}~\bibnamefont {Tsukamoto}}, \bibinfo
  {author} {\bibfnamefont {A.}~\bibnamefont {Fognini}}, \bibinfo {author}
  {\bibfnamefont {J.~J.}\ \bibnamefont {Turner}}, \bibinfo {author}
  {\bibfnamefont {W.~F.}\ \bibnamefont {Schlotter}}, \bibinfo {author}
  {\bibfnamefont {D.}~\bibnamefont {Rolles}}, \bibinfo {author} {\bibfnamefont
  {H.}~\bibnamefont {Soltau}}, \bibinfo {author} {\bibfnamefont
  {L.}~\bibnamefont {Str{\"u}der}}, \bibinfo {author} {\bibfnamefont
  {Y.}~\bibnamefont {Acremann}}, \bibinfo {author} {\bibfnamefont {A.~V.}\
  \bibnamefont {Kimel}}, \bibinfo {author} {\bibfnamefont {A.}~\bibnamefont
  {Kirilyuk}}, \bibinfo {author} {\bibfnamefont {T.}~\bibnamefont {Rasing}},
  \bibinfo {author} {\bibfnamefont {J.}~\bibnamefont {St{\"o}hr}}, \bibinfo
  {author} {\bibfnamefont {A.~O.}\ \bibnamefont {Scherz}}, \ and\ \bibinfo
  {author} {\bibfnamefont {H.~A.}\ \bibnamefont {D{\"u}rr}},\ }\href {\doibase
  10.1038/nmat3597} {\bibfield  {journal} {\bibinfo  {journal} {Nature
  Materials}\ }\textbf {\bibinfo {volume} {12}},\ \bibinfo {pages} {293}
  (\bibinfo {year} {2013})}\BibitemShut {NoStop}%
\bibitem [{\citenamefont {Wieczorek}\ \emph {et~al.}(2015)\citenamefont
  {Wieczorek}, \citenamefont {Eschenlohr}, \citenamefont {Weidtmann},
  \citenamefont {R\"osner}, \citenamefont {Bergeard}, \citenamefont
  {Tarasevitch}, \citenamefont {Wehling},\ and\ \citenamefont
  {Bovensiepen}}]{Wieczorek2015}%
  \BibitemOpen
  \bibfield  {author} {\bibinfo {author} {\bibfnamefont {J.}~\bibnamefont
  {Wieczorek}}, \bibinfo {author} {\bibfnamefont {A.}~\bibnamefont
  {Eschenlohr}}, \bibinfo {author} {\bibfnamefont {B.}~\bibnamefont
  {Weidtmann}}, \bibinfo {author} {\bibfnamefont {M.}~\bibnamefont {R\"osner}},
  \bibinfo {author} {\bibfnamefont {N.}~\bibnamefont {Bergeard}}, \bibinfo
  {author} {\bibfnamefont {A.}~\bibnamefont {Tarasevitch}}, \bibinfo {author}
  {\bibfnamefont {T.~O.}\ \bibnamefont {Wehling}}, \ and\ \bibinfo {author}
  {\bibfnamefont {U.}~\bibnamefont {Bovensiepen}},\ }\href {\doibase
  10.1103/PhysRevB.92.174410} {\bibfield  {journal} {\bibinfo  {journal} {Phys.
  Rev. B}\ }\textbf {\bibinfo {volume} {92}},\ \bibinfo {pages} {174410}
  (\bibinfo {year} {2015})}\BibitemShut {NoStop}%
\bibitem [{\citenamefont {Elyasi}\ and\ \citenamefont
  {Yang}(2016)}]{Elyasi2016}%
  \BibitemOpen
  \bibfield  {author} {\bibinfo {author} {\bibfnamefont {M.}~\bibnamefont
  {Elyasi}}\ and\ \bibinfo {author} {\bibfnamefont {H.}~\bibnamefont {Yang}},\
  }\href {\doibase 10.1103/PhysRevB.94.024417} {\bibfield  {journal} {\bibinfo
  {journal} {Phys. Rev. B}\ }\textbf {\bibinfo {volume} {94}},\ \bibinfo
  {pages} {024417} (\bibinfo {year} {2016})}\BibitemShut {NoStop}%
\bibitem [{\citenamefont {Tengdin}\ \emph {et~al.}(2018)\citenamefont
  {Tengdin}, \citenamefont {You}, \citenamefont {Chen}, \citenamefont {Shi},
  \citenamefont {Zusin}, \citenamefont {Zhang}, \citenamefont {Gentry},
  \citenamefont {Blonsky}, \citenamefont {Keller}, \citenamefont {Oppeneer},
  \citenamefont {Kapteyn}, \citenamefont {Tao},\ and\ \citenamefont
  {Murnane}}]{Tengdineaap2018}%
  \BibitemOpen
  \bibfield  {author} {\bibinfo {author} {\bibfnamefont {P.}~\bibnamefont
  {Tengdin}}, \bibinfo {author} {\bibfnamefont {W.}~\bibnamefont {You}},
  \bibinfo {author} {\bibfnamefont {C.}~\bibnamefont {Chen}}, \bibinfo {author}
  {\bibfnamefont {X.}~\bibnamefont {Shi}}, \bibinfo {author} {\bibfnamefont
  {D.}~\bibnamefont {Zusin}}, \bibinfo {author} {\bibfnamefont
  {Y.}~\bibnamefont {Zhang}}, \bibinfo {author} {\bibfnamefont
  {C.}~\bibnamefont {Gentry}}, \bibinfo {author} {\bibfnamefont
  {A.}~\bibnamefont {Blonsky}}, \bibinfo {author} {\bibfnamefont
  {M.}~\bibnamefont {Keller}}, \bibinfo {author} {\bibfnamefont {P.~M.}\
  \bibnamefont {Oppeneer}}, \bibinfo {author} {\bibfnamefont {H.~C.}\
  \bibnamefont {Kapteyn}}, \bibinfo {author} {\bibfnamefont {Z.}~\bibnamefont
  {Tao}}, \ and\ \bibinfo {author} {\bibfnamefont {M.~M.}\ \bibnamefont
  {Murnane}},\ }\href {https://advances.sciencemag.org/content/4/3/eaap9744}
  {\bibfield  {journal} {\bibinfo  {journal} {Science Advances}\ }\textbf
  {\bibinfo {volume} {4}} (\bibinfo {year} {2018})}\BibitemShut {NoStop}%
\bibitem [{\citenamefont {Schneider}\ \emph {et~al.}(2018)\citenamefont
  {Schneider}, \citenamefont {G{\"u}nther}, \citenamefont {Pfau}, \citenamefont
  {Capotondi}, \citenamefont {Manfredda}, \citenamefont {Zangrando},
  \citenamefont {Mahne}, \citenamefont {Raimondi}, \citenamefont {Pedersoli},
  \citenamefont {Naumenko},\ and\ \citenamefont {Eisebitt}}]{Schneider2018}%
  \BibitemOpen
  \bibfield  {author} {\bibinfo {author} {\bibfnamefont {M.}~\bibnamefont
  {Schneider}}, \bibinfo {author} {\bibfnamefont {C.~M.}\ \bibnamefont
  {G{\"u}nther}}, \bibinfo {author} {\bibfnamefont {B.}~\bibnamefont {Pfau}},
  \bibinfo {author} {\bibfnamefont {F.}~\bibnamefont {Capotondi}}, \bibinfo
  {author} {\bibfnamefont {M.}~\bibnamefont {Manfredda}}, \bibinfo {author}
  {\bibfnamefont {M.}~\bibnamefont {Zangrando}}, \bibinfo {author}
  {\bibfnamefont {N.}~\bibnamefont {Mahne}}, \bibinfo {author} {\bibfnamefont
  {L.}~\bibnamefont {Raimondi}}, \bibinfo {author} {\bibfnamefont
  {E.}~\bibnamefont {Pedersoli}}, \bibinfo {author} {\bibfnamefont
  {D.}~\bibnamefont {Naumenko}}, \ and\ \bibinfo {author} {\bibfnamefont
  {S.}~\bibnamefont {Eisebitt}},\ }\href {\doibase 10.1038/s41467-017-02567-0}
  {\bibfield  {journal} {\bibinfo  {journal} {Nature Communications}\ }\textbf
  {\bibinfo {volume} {9}},\ \bibinfo {pages} {214} (\bibinfo {year}
  {2018})}\BibitemShut {NoStop}%
\bibitem [{\citenamefont {Kortright}\ \emph {et~al.}(2001)\citenamefont
  {Kortright}, \citenamefont {Kim}, \citenamefont {Denbeaux}, \citenamefont
  {Zeltzer}, \citenamefont {Takano},\ and\ \citenamefont
  {Fullerton}}]{Kortright2001}%
  \BibitemOpen
  \bibfield  {author} {\bibinfo {author} {\bibfnamefont {J.~B.}\ \bibnamefont
  {Kortright}}, \bibinfo {author} {\bibfnamefont {S.-K.}\ \bibnamefont {Kim}},
  \bibinfo {author} {\bibfnamefont {G.~P.}\ \bibnamefont {Denbeaux}}, \bibinfo
  {author} {\bibfnamefont {G.}~\bibnamefont {Zeltzer}}, \bibinfo {author}
  {\bibfnamefont {K.}~\bibnamefont {Takano}}, \ and\ \bibinfo {author}
  {\bibfnamefont {E.~E.}\ \bibnamefont {Fullerton}},\ }\href {\doibase
  10.1103/PhysRevB.64.092401} {\bibfield  {journal} {\bibinfo  {journal} {Phys.
  Rev. B}\ }\textbf {\bibinfo {volume} {64}},\ \bibinfo {pages} {092401}
  (\bibinfo {year} {2001})}\BibitemShut {NoStop}%
\bibitem [{\citenamefont {Hellwig}\ \emph {et~al.}(2003)\citenamefont
  {Hellwig}, \citenamefont {Denbeaux}, \citenamefont {Kortright},\ and\
  \citenamefont {Fullerton}}]{Hellwig2003}%
  \BibitemOpen
  \bibfield  {author} {\bibinfo {author} {\bibfnamefont {O.}~\bibnamefont
  {Hellwig}}, \bibinfo {author} {\bibfnamefont {G.}~\bibnamefont {Denbeaux}},
  \bibinfo {author} {\bibfnamefont {J.}~\bibnamefont {Kortright}}, \ and\
  \bibinfo {author} {\bibfnamefont {E.~E.}\ \bibnamefont {Fullerton}},\ }\href
  {\doibase https://doi.org/10.1016/S0921-4526(03)00282-5} {\bibfield
  {journal} {\bibinfo  {journal} {Physica B: Condensed Matter}\ }\textbf
  {\bibinfo {volume} {336}},\ \bibinfo {pages} {136 } (\bibinfo {year}
  {2003})},\ \bibinfo {note} {{P}roceedings of the Seventh International
  Conference on Surface X-ray and Neutron Scattering}\BibitemShut {NoStop}%
\bibitem [{\citenamefont {Zusin}\ \emph {et~al.}(2020)\citenamefont {Zusin},
  \citenamefont {Iacocca}, \citenamefont {Guyader}, \citenamefont {Reid},
  \citenamefont {Schlotter}, \citenamefont {Liu}, \citenamefont {Higley},
  \citenamefont {Coslovich}, \citenamefont {Wandel}, \citenamefont {Tengdin},
  \citenamefont {Patel}, \citenamefont {Shabalin}, \citenamefont {Hua},
  \citenamefont {Hrkac}, \citenamefont {Nembach}, \citenamefont {Shaw},
  \citenamefont {Montoya}, \citenamefont {Blonsky}, \citenamefont {Gentry},
  \citenamefont {Hoefer}, \citenamefont {Murnane}, \citenamefont {Kapteyn},
  \citenamefont {Fullerton}, \citenamefont {Shpyrko}, \citenamefont {Dürr},\
  and\ \citenamefont {Silva}}]{Zusin2020}%
  \BibitemOpen
  \bibfield  {author} {\bibinfo {author} {\bibfnamefont {D.}~\bibnamefont
  {Zusin}}, \bibinfo {author} {\bibfnamefont {E.}~\bibnamefont {Iacocca}},
  \bibinfo {author} {\bibfnamefont {L.~L.}\ \bibnamefont {Guyader}}, \bibinfo
  {author} {\bibfnamefont {A.~H.}\ \bibnamefont {Reid}}, \bibinfo {author}
  {\bibfnamefont {W.~F.}\ \bibnamefont {Schlotter}}, \bibinfo {author}
  {\bibfnamefont {T.-M.}\ \bibnamefont {Liu}}, \bibinfo {author} {\bibfnamefont
  {D.~J.}\ \bibnamefont {Higley}}, \bibinfo {author} {\bibfnamefont
  {G.}~\bibnamefont {Coslovich}}, \bibinfo {author} {\bibfnamefont {S.~F.}\
  \bibnamefont {Wandel}}, \bibinfo {author} {\bibfnamefont {P.~M.}\
  \bibnamefont {Tengdin}}, \bibinfo {author} {\bibfnamefont {S.~K.~K.}\
  \bibnamefont {Patel}}, \bibinfo {author} {\bibfnamefont {A.}~\bibnamefont
  {Shabalin}}, \bibinfo {author} {\bibfnamefont {N.}~\bibnamefont {Hua}},
  \bibinfo {author} {\bibfnamefont {S.~B.}\ \bibnamefont {Hrkac}}, \bibinfo
  {author} {\bibfnamefont {H.~T.}\ \bibnamefont {Nembach}}, \bibinfo {author}
  {\bibfnamefont {J.~M.}\ \bibnamefont {Shaw}}, \bibinfo {author}
  {\bibfnamefont {S.~A.}\ \bibnamefont {Montoya}}, \bibinfo {author}
  {\bibfnamefont {A.}~\bibnamefont {Blonsky}}, \bibinfo {author} {\bibfnamefont
  {C.}~\bibnamefont {Gentry}}, \bibinfo {author} {\bibfnamefont {M.~A.}\
  \bibnamefont {Hoefer}}, \bibinfo {author} {\bibfnamefont {M.~M.}\
  \bibnamefont {Murnane}}, \bibinfo {author} {\bibfnamefont {H.~C.}\
  \bibnamefont {Kapteyn}}, \bibinfo {author} {\bibfnamefont {E.~E.}\
  \bibnamefont {Fullerton}}, \bibinfo {author} {\bibfnamefont {O.}~\bibnamefont
  {Shpyrko}}, \bibinfo {author} {\bibfnamefont {H.~A.}\ \bibnamefont {Dürr}},
  \ and\ \bibinfo {author} {\bibfnamefont {T.~J.}\ \bibnamefont {Silva}},\
  }\href@noop {} {\enquote {\bibinfo {title} {Ultrafast domain dilation induced
  by optical pumping in ferromagnetic {C}o{F}e/{N}i multilayers},}\ } (\bibinfo
  {year} {2020}),\ \Eprint {http://arxiv.org/abs/2001.11719} {arXiv:2001.11719
  [cond-mat.mes-hall]} \BibitemShut {NoStop}%
\bibitem [{\citenamefont {Moisan}\ \emph
  {et~al.}(2014{\natexlab{a}})\citenamefont {Moisan}, \citenamefont
  {Malinowski}, \citenamefont {Mauchain}, \citenamefont {Hehn}, \citenamefont
  {Vodungbo}, \citenamefont {L{\"u}ning}, \citenamefont {Mangin}, \citenamefont
  {Fullerton},\ and\ \citenamefont {Thiaville}}]{Moisan2014}%
  \BibitemOpen
  \bibfield  {author} {\bibinfo {author} {\bibfnamefont {N.}~\bibnamefont
  {Moisan}}, \bibinfo {author} {\bibfnamefont {G.}~\bibnamefont {Malinowski}},
  \bibinfo {author} {\bibfnamefont {J.}~\bibnamefont {Mauchain}}, \bibinfo
  {author} {\bibfnamefont {M.}~\bibnamefont {Hehn}}, \bibinfo {author}
  {\bibfnamefont {B.}~\bibnamefont {Vodungbo}}, \bibinfo {author}
  {\bibfnamefont {J.}~\bibnamefont {L{\"u}ning}}, \bibinfo {author}
  {\bibfnamefont {S.}~\bibnamefont {Mangin}}, \bibinfo {author} {\bibfnamefont
  {E.~E.}\ \bibnamefont {Fullerton}}, \ and\ \bibinfo {author} {\bibfnamefont
  {A.}~\bibnamefont {Thiaville}},\ }\href {\doibase 10.1038/srep04658}
  {\bibfield  {journal} {\bibinfo  {journal} {Scientific Reports}\ }\textbf
  {\bibinfo {volume} {4}},\ \bibinfo {pages} {4658} (\bibinfo {year}
  {2014}{\natexlab{a}})}\BibitemShut {NoStop}%
\bibitem [{\citenamefont {L\'opez-Flores}\ \emph {et~al.}(2013)\citenamefont
  {L\'opez-Flores}, \citenamefont {Bergeard}, \citenamefont {Halt\'e},
  \citenamefont {Stamm}, \citenamefont {Pontius}, \citenamefont {Hehn},
  \citenamefont {Otero}, \citenamefont {Beaurepaire},\ and\ \citenamefont
  {Boeglin}}]{LopezFlores2013}%
  \BibitemOpen
  \bibfield  {author} {\bibinfo {author} {\bibfnamefont {V.}~\bibnamefont
  {L\'opez-Flores}}, \bibinfo {author} {\bibfnamefont {N.}~\bibnamefont
  {Bergeard}}, \bibinfo {author} {\bibfnamefont {V.}~\bibnamefont {Halt\'e}},
  \bibinfo {author} {\bibfnamefont {C.}~\bibnamefont {Stamm}}, \bibinfo
  {author} {\bibfnamefont {N.}~\bibnamefont {Pontius}}, \bibinfo {author}
  {\bibfnamefont {M.}~\bibnamefont {Hehn}}, \bibinfo {author} {\bibfnamefont
  {E.}~\bibnamefont {Otero}}, \bibinfo {author} {\bibfnamefont
  {E.}~\bibnamefont {Beaurepaire}}, \ and\ \bibinfo {author} {\bibfnamefont
  {C.}~\bibnamefont {Boeglin}},\ }\href {\doibase 10.1103/PhysRevB.87.214412}
  {\bibfield  {journal} {\bibinfo  {journal} {Phys. Rev. B}\ }\textbf {\bibinfo
  {volume} {87}},\ \bibinfo {pages} {214412} (\bibinfo {year}
  {2013})}\BibitemShut {NoStop}%
\bibitem [{\citenamefont {Bergeard}\ \emph {et~al.}(2014)\citenamefont
  {Bergeard}, \citenamefont {L{\'o}pez-Flores}, \citenamefont {Halt{\'e}},
  \citenamefont {Hehn}, \citenamefont {Stamm}, \citenamefont {Pontius},
  \citenamefont {Beaurepaire},\ and\ \citenamefont {Boeglin}}]{Bergeard2014}%
  \BibitemOpen
  \bibfield  {author} {\bibinfo {author} {\bibfnamefont {N.}~\bibnamefont
  {Bergeard}}, \bibinfo {author} {\bibfnamefont {V.}~\bibnamefont
  {L{\'o}pez-Flores}}, \bibinfo {author} {\bibfnamefont {V.}~\bibnamefont
  {Halt{\'e}}}, \bibinfo {author} {\bibfnamefont {M.}~\bibnamefont {Hehn}},
  \bibinfo {author} {\bibfnamefont {C.}~\bibnamefont {Stamm}}, \bibinfo
  {author} {\bibfnamefont {N.}~\bibnamefont {Pontius}}, \bibinfo {author}
  {\bibfnamefont {E.}~\bibnamefont {Beaurepaire}}, \ and\ \bibinfo {author}
  {\bibfnamefont {C.}~\bibnamefont {Boeglin}},\ }\href {\doibase
  10.1038/ncomms4466} {\bibfield  {journal} {\bibinfo  {journal} {Nature
  Communications}\ }\textbf {\bibinfo {volume} {5}},\ \bibinfo {pages} {3466}
  (\bibinfo {year} {2014})}\BibitemShut {NoStop}%
\bibitem [{\citenamefont {Alebrand}\ \emph {et~al.}(2014)\citenamefont
  {Alebrand}, \citenamefont {Bierbrauer}, \citenamefont {Hehn}, \citenamefont
  {Gottwald}, \citenamefont {Schmitt}, \citenamefont {Steil}, \citenamefont
  {Fullerton}, \citenamefont {Mangin}, \citenamefont {Cinchetti},\ and\
  \citenamefont {Aeschlimann}}]{Alebrand2014}%
  \BibitemOpen
  \bibfield  {author} {\bibinfo {author} {\bibfnamefont {S.}~\bibnamefont
  {Alebrand}}, \bibinfo {author} {\bibfnamefont {U.}~\bibnamefont
  {Bierbrauer}}, \bibinfo {author} {\bibfnamefont {M.}~\bibnamefont {Hehn}},
  \bibinfo {author} {\bibfnamefont {M.}~\bibnamefont {Gottwald}}, \bibinfo
  {author} {\bibfnamefont {O.}~\bibnamefont {Schmitt}}, \bibinfo {author}
  {\bibfnamefont {D.}~\bibnamefont {Steil}}, \bibinfo {author} {\bibfnamefont
  {E.~E.}\ \bibnamefont {Fullerton}}, \bibinfo {author} {\bibfnamefont
  {S.}~\bibnamefont {Mangin}}, \bibinfo {author} {\bibfnamefont
  {M.}~\bibnamefont {Cinchetti}}, \ and\ \bibinfo {author} {\bibfnamefont
  {M.}~\bibnamefont {Aeschlimann}},\ }\href {\doibase
  10.1103/PhysRevB.89.144404} {\bibfield  {journal} {\bibinfo  {journal} {Phys.
  Rev. B}\ }\textbf {\bibinfo {volume} {89}},\ \bibinfo {pages} {144404}
  (\bibinfo {year} {2014})}\BibitemShut {NoStop}%
\bibitem [{\citenamefont {Fert\'e}\ \emph {et~al.}(2017)\citenamefont
  {Fert\'e}, \citenamefont {Bergeard}, \citenamefont {Malinowski},
  \citenamefont {Abrudan}, \citenamefont {Kachel}, \citenamefont {Holldack},
  \citenamefont {Hehn},\ and\ \citenamefont {Boeglin}}]{Ferte2017}%
  \BibitemOpen
  \bibfield  {author} {\bibinfo {author} {\bibfnamefont {T.}~\bibnamefont
  {Fert\'e}}, \bibinfo {author} {\bibfnamefont {N.}~\bibnamefont {Bergeard}},
  \bibinfo {author} {\bibfnamefont {G.}~\bibnamefont {Malinowski}}, \bibinfo
  {author} {\bibfnamefont {R.}~\bibnamefont {Abrudan}}, \bibinfo {author}
  {\bibfnamefont {T.}~\bibnamefont {Kachel}}, \bibinfo {author} {\bibfnamefont
  {K.}~\bibnamefont {Holldack}}, \bibinfo {author} {\bibfnamefont
  {M.}~\bibnamefont {Hehn}}, \ and\ \bibinfo {author} {\bibfnamefont
  {C.}~\bibnamefont {Boeglin}},\ }\href {\doibase 10.1103/PhysRevB.96.144427}
  {\bibfield  {journal} {\bibinfo  {journal} {Phys. Rev. B}\ }\textbf {\bibinfo
  {volume} {96}},\ \bibinfo {pages} {144427} (\bibinfo {year}
  {2017})}\BibitemShut {NoStop}%
\bibitem [{\citenamefont {Stanciu}\ \emph {et~al.}(2007)\citenamefont
  {Stanciu}, \citenamefont {Hansteen}, \citenamefont {Kimel}, \citenamefont
  {Kirilyuk}, \citenamefont {Tsukamoto}, \citenamefont {Itoh},\ and\
  \citenamefont {Rasing}}]{Stanciu2007}%
  \BibitemOpen
  \bibfield  {author} {\bibinfo {author} {\bibfnamefont {C.~D.}\ \bibnamefont
  {Stanciu}}, \bibinfo {author} {\bibfnamefont {F.}~\bibnamefont {Hansteen}},
  \bibinfo {author} {\bibfnamefont {A.~V.}\ \bibnamefont {Kimel}}, \bibinfo
  {author} {\bibfnamefont {A.}~\bibnamefont {Kirilyuk}}, \bibinfo {author}
  {\bibfnamefont {A.}~\bibnamefont {Tsukamoto}}, \bibinfo {author}
  {\bibfnamefont {A.}~\bibnamefont {Itoh}}, \ and\ \bibinfo {author}
  {\bibfnamefont {T.}~\bibnamefont {Rasing}},\ }\href {\doibase
  10.1103/PhysRevLett.99.047601} {\bibfield  {journal} {\bibinfo  {journal}
  {Phys. Rev. Lett.}\ }\textbf {\bibinfo {volume} {99}},\ \bibinfo {pages}
  {047601} (\bibinfo {year} {2007})}\BibitemShut {NoStop}%
\bibitem [{\citenamefont {Alebrand}\ \emph {et~al.}(2012)\citenamefont
  {Alebrand}, \citenamefont {Gottwald}, \citenamefont {Hehn}, \citenamefont
  {Steil}, \citenamefont {Cinchetti}, \citenamefont {Lacour}, \citenamefont
  {Fullerton}, \citenamefont {Aeschlimann},\ and\ \citenamefont
  {Mangin}}]{Alebrand2012}%
  \BibitemOpen
  \bibfield  {author} {\bibinfo {author} {\bibfnamefont {S.}~\bibnamefont
  {Alebrand}}, \bibinfo {author} {\bibfnamefont {M.}~\bibnamefont {Gottwald}},
  \bibinfo {author} {\bibfnamefont {M.}~\bibnamefont {Hehn}}, \bibinfo {author}
  {\bibfnamefont {D.}~\bibnamefont {Steil}}, \bibinfo {author} {\bibfnamefont
  {M.}~\bibnamefont {Cinchetti}}, \bibinfo {author} {\bibfnamefont
  {D.}~\bibnamefont {Lacour}}, \bibinfo {author} {\bibfnamefont {E.~E.}\
  \bibnamefont {Fullerton}}, \bibinfo {author} {\bibfnamefont {M.}~\bibnamefont
  {Aeschlimann}}, \ and\ \bibinfo {author} {\bibfnamefont {S.}~\bibnamefont
  {Mangin}},\ }\href {https://doi.org/10.1063/1.4759109} {\bibfield  {journal}
  {\bibinfo  {journal} {Applied Physics Letters}\ }\textbf {\bibinfo {volume}
  {101}},\ \bibinfo {pages} {162408} (\bibinfo {year} {2012})}\BibitemShut
  {NoStop}%
\bibitem [{\citenamefont {El~Hadri}\ \emph
  {et~al.}(2016{\natexlab{a}})\citenamefont {El~Hadri}, \citenamefont {Hehn},
  \citenamefont {Pirro}, \citenamefont {Lambert}, \citenamefont {Malinowski},
  \citenamefont {Fullerton},\ and\ \citenamefont {Mangin}}]{Hadri2016}%
  \BibitemOpen
  \bibfield  {author} {\bibinfo {author} {\bibfnamefont {M.~S.}\ \bibnamefont
  {El~Hadri}}, \bibinfo {author} {\bibfnamefont {M.}~\bibnamefont {Hehn}},
  \bibinfo {author} {\bibfnamefont {P.}~\bibnamefont {Pirro}}, \bibinfo
  {author} {\bibfnamefont {C.-H.}\ \bibnamefont {Lambert}}, \bibinfo {author}
  {\bibfnamefont {G.}~\bibnamefont {Malinowski}}, \bibinfo {author}
  {\bibfnamefont {E.~E.}\ \bibnamefont {Fullerton}}, \ and\ \bibinfo {author}
  {\bibfnamefont {S.}~\bibnamefont {Mangin}},\ }\href {\doibase
  10.1103/PhysRevB.94.064419} {\bibfield  {journal} {\bibinfo  {journal} {Phys.
  Rev. B}\ }\textbf {\bibinfo {volume} {94}},\ \bibinfo {pages} {064419}
  (\bibinfo {year} {2016}{\natexlab{a}})}\BibitemShut {NoStop}%
\bibitem [{\citenamefont {Fan}\ \emph {et~al.}(2019)\citenamefont {Fan},
  \citenamefont {Legare}, \citenamefont {Cardin}, \citenamefont {Xie},
  \citenamefont {Kaksis}, \citenamefont {Andriukaitis}, \citenamefont
  {Pugzlys}, \citenamefont {Schmidt}, \citenamefont {Wolf}, \citenamefont
  {Hehn}, \citenamefont {Malinowski}, \citenamefont {Vodungbo}, \citenamefont
  {Jal}, \citenamefont {Luning}, \citenamefont {Jaouen}, \citenamefont {Tao},
  \citenamefont {Baltuska}, \citenamefont {Legare},\ and\ \citenamefont
  {Balciunas}}]{Fan2019}%
  \BibitemOpen
  \bibfield  {author} {\bibinfo {author} {\bibfnamefont {G.}~\bibnamefont
  {Fan}}, \bibinfo {author} {\bibfnamefont {K.}~\bibnamefont {Legare}},
  \bibinfo {author} {\bibfnamefont {V.}~\bibnamefont {Cardin}}, \bibinfo
  {author} {\bibfnamefont {X.}~\bibnamefont {Xie}}, \bibinfo {author}
  {\bibfnamefont {E.}~\bibnamefont {Kaksis}}, \bibinfo {author} {\bibfnamefont
  {G.}~\bibnamefont {Andriukaitis}}, \bibinfo {author} {\bibfnamefont
  {A.}~\bibnamefont {Pugzlys}}, \bibinfo {author} {\bibfnamefont {B.~E.}\
  \bibnamefont {Schmidt}}, \bibinfo {author} {\bibfnamefont {J.~P.}\
  \bibnamefont {Wolf}}, \bibinfo {author} {\bibfnamefont {M.}~\bibnamefont
  {Hehn}}, \bibinfo {author} {\bibfnamefont {G.}~\bibnamefont {Malinowski}},
  \bibinfo {author} {\bibfnamefont {B.}~\bibnamefont {Vodungbo}}, \bibinfo
  {author} {\bibfnamefont {E.}~\bibnamefont {Jal}}, \bibinfo {author}
  {\bibfnamefont {J.}~\bibnamefont {Luning}}, \bibinfo {author} {\bibfnamefont
  {N.}~\bibnamefont {Jaouen}}, \bibinfo {author} {\bibfnamefont
  {Z.}~\bibnamefont {Tao}}, \bibinfo {author} {\bibfnamefont {A.}~\bibnamefont
  {Baltuska}}, \bibinfo {author} {\bibfnamefont {F.}~\bibnamefont {Legare}}, \
  and\ \bibinfo {author} {\bibfnamefont {T.}~\bibnamefont {Balciunas}},\
  }\href@noop {} {\enquote {\bibinfo {title} {Time-resolving magnetic
  scattering on rare-earth ferrimagnets with a bright soft-x-ray high-harmonic
  source},}\ } (\bibinfo {year} {2019}),\ \Eprint
  {http://arxiv.org/abs/1910.14263} {arXiv:1910.14263 [physics.optics]}
  \BibitemShut {NoStop}%
\bibitem [{\citenamefont {Gottwald}\ \emph {et~al.}(2012)\citenamefont
  {Gottwald}, \citenamefont {Hehn}, \citenamefont {Montaigne}, \citenamefont
  {Lacour}, \citenamefont {Lengaigne}, \citenamefont {Suire},\ and\
  \citenamefont {Mangin}}]{Gottwald2012}%
  \BibitemOpen
  \bibfield  {author} {\bibinfo {author} {\bibfnamefont {M.}~\bibnamefont
  {Gottwald}}, \bibinfo {author} {\bibfnamefont {M.}~\bibnamefont {Hehn}},
  \bibinfo {author} {\bibfnamefont {F.}~\bibnamefont {Montaigne}}, \bibinfo
  {author} {\bibfnamefont {D.}~\bibnamefont {Lacour}}, \bibinfo {author}
  {\bibfnamefont {G.}~\bibnamefont {Lengaigne}}, \bibinfo {author}
  {\bibfnamefont {S.}~\bibnamefont {Suire}}, \ and\ \bibinfo {author}
  {\bibfnamefont {S.}~\bibnamefont {Mangin}},\ }\href
  {https://doi.org/10.1063/1.3703666} {\bibfield  {journal} {\bibinfo
  {journal} {Journal of Applied Physics}\ }\textbf {\bibinfo {volume} {111}},\
  \bibinfo {pages} {083904} (\bibinfo {year} {2012})}\BibitemShut {NoStop}%
\bibitem [{\citenamefont {Capotondi}\ \emph {et~al.}(2013)\citenamefont
  {Capotondi}, \citenamefont {Pedersoli}, \citenamefont {Mahne}, \citenamefont
  {Menk}, \citenamefont {Passos}, \citenamefont {Raimondi}, \citenamefont
  {Svetina}, \citenamefont {Sandrin}, \citenamefont {Zangrando}, \citenamefont
  {Kiskinova}, \citenamefont {Bajt}, \citenamefont {Barthelmess}, \citenamefont
  {Fleckenstein}, \citenamefont {Chapman}, \citenamefont {Schulz},
  \citenamefont {Bach}, \citenamefont {Fr\"{o}mter}, \citenamefont
  {Schleitzer}, \citenamefont {M\"{u}ller}, \citenamefont {Gutt},\ and\
  \citenamefont {Gr\"{u}bel}}]{Capotondi2013}%
  \BibitemOpen
  \bibfield  {author} {\bibinfo {author} {\bibfnamefont {F.}~\bibnamefont
  {Capotondi}}, \bibinfo {author} {\bibfnamefont {E.}~\bibnamefont
  {Pedersoli}}, \bibinfo {author} {\bibfnamefont {N.}~\bibnamefont {Mahne}},
  \bibinfo {author} {\bibfnamefont {R.~H.}\ \bibnamefont {Menk}}, \bibinfo
  {author} {\bibfnamefont {G.}~\bibnamefont {Passos}}, \bibinfo {author}
  {\bibfnamefont {L.}~\bibnamefont {Raimondi}}, \bibinfo {author}
  {\bibfnamefont {C.}~\bibnamefont {Svetina}}, \bibinfo {author} {\bibfnamefont
  {G.}~\bibnamefont {Sandrin}}, \bibinfo {author} {\bibfnamefont
  {M.}~\bibnamefont {Zangrando}}, \bibinfo {author} {\bibfnamefont
  {M.}~\bibnamefont {Kiskinova}}, \bibinfo {author} {\bibfnamefont
  {S.}~\bibnamefont {Bajt}}, \bibinfo {author} {\bibfnamefont {M.}~\bibnamefont
  {Barthelmess}}, \bibinfo {author} {\bibfnamefont {H.}~\bibnamefont
  {Fleckenstein}}, \bibinfo {author} {\bibfnamefont {H.~N.}\ \bibnamefont
  {Chapman}}, \bibinfo {author} {\bibfnamefont {J.}~\bibnamefont {Schulz}},
  \bibinfo {author} {\bibfnamefont {J.}~\bibnamefont {Bach}}, \bibinfo {author}
  {\bibfnamefont {R.}~\bibnamefont {Fr\"{o}mter}}, \bibinfo {author}
  {\bibfnamefont {S.}~\bibnamefont {Schleitzer}}, \bibinfo {author}
  {\bibfnamefont {L.}~\bibnamefont {M\"{u}ller}}, \bibinfo {author}
  {\bibfnamefont {C.}~\bibnamefont {Gutt}}, \ and\ \bibinfo {author}
  {\bibfnamefont {G.}~\bibnamefont {Gr\"{u}bel}},\ }\href
  {https://doi.org/10.1063/1.4807157} {\bibfield  {journal} {\bibinfo
  {journal} {Review of Scientific Instruments}\ }\textbf {\bibinfo {volume}
  {84}},\ \bibinfo {pages} {051301} (\bibinfo {year} {2013})}\BibitemShut
  {NoStop}%
\bibitem [{\citenamefont {Harris}\ \emph {et~al.}(1992)\citenamefont {Harris},
  \citenamefont {Aylesworth}, \citenamefont {Das}, \citenamefont {Elam},\ and\
  \citenamefont {Koon}}]{Harris1992}%
  \BibitemOpen
  \bibfield  {author} {\bibinfo {author} {\bibfnamefont {V.~G.}\ \bibnamefont
  {Harris}}, \bibinfo {author} {\bibfnamefont {K.~D.}\ \bibnamefont
  {Aylesworth}}, \bibinfo {author} {\bibfnamefont {B.~N.}\ \bibnamefont {Das}},
  \bibinfo {author} {\bibfnamefont {W.~T.}\ \bibnamefont {Elam}}, \ and\
  \bibinfo {author} {\bibfnamefont {N.~C.}\ \bibnamefont {Koon}},\ }\href
  {\doibase 10.1103/PhysRevLett.69.1939} {\bibfield  {journal} {\bibinfo
  {journal} {Phys. Rev. Lett.}\ }\textbf {\bibinfo {volume} {69}},\ \bibinfo
  {pages} {1939} (\bibinfo {year} {1992})}\BibitemShut {NoStop}%
\bibitem [{\citenamefont {Kittel}(1949)}]{Kittel1949}%
  \BibitemOpen
  \bibfield  {author} {\bibinfo {author} {\bibfnamefont {C.}~\bibnamefont
  {Kittel}},\ }\href {\doibase 10.1103/RevModPhys.21.541} {\bibfield  {journal}
  {\bibinfo  {journal} {Rev. Mod. Phys.}\ }\textbf {\bibinfo {volume} {21}},\
  \bibinfo {pages} {541} (\bibinfo {year} {1949})}\BibitemShut {NoStop}%
\bibitem [{\citenamefont {Chen}\ \emph {et~al.}(2019)\citenamefont {Chen},
  \citenamefont {Li}, \citenamefont {Zhou},\ and\ \citenamefont
  {Lai}}]{Lai2019}%
  \BibitemOpen
  \bibfield  {author} {\bibinfo {author} {\bibfnamefont {Z.}~\bibnamefont
  {Chen}}, \bibinfo {author} {\bibfnamefont {S.}~\bibnamefont {Li}}, \bibinfo
  {author} {\bibfnamefont {S.}~\bibnamefont {Zhou}}, \ and\ \bibinfo {author}
  {\bibfnamefont {T.}~\bibnamefont {Lai}},\ }\href {\doibase
  10.1088/1367-2630/ab5aa4} {\bibfield  {journal} {\bibinfo  {journal} {New
  Journal of Physics}\ }\textbf {\bibinfo {volume} {21}},\ \bibinfo {pages}
  {123007} (\bibinfo {year} {2019})}\BibitemShut {NoStop}%
\bibitem [{\citenamefont {Radu}\ \emph {et~al.}(2011)\citenamefont {Radu},
  \citenamefont {Vahaplar}, \citenamefont {Stamm}, \citenamefont {Kachel},
  \citenamefont {Pontius}, \citenamefont {D{\"u}rr}, \citenamefont {Ostler},
  \citenamefont {Barker}, \citenamefont {Evans}, \citenamefont {Chantrell},
  \citenamefont {Tsukamoto}, \citenamefont {Itoh}, \citenamefont {Kirilyuk},
  \citenamefont {Rasing},\ and\ \citenamefont {Kimel}}]{Radu2011}%
  \BibitemOpen
  \bibfield  {author} {\bibinfo {author} {\bibfnamefont {I.}~\bibnamefont
  {Radu}}, \bibinfo {author} {\bibfnamefont {K.}~\bibnamefont {Vahaplar}},
  \bibinfo {author} {\bibfnamefont {C.}~\bibnamefont {Stamm}}, \bibinfo
  {author} {\bibfnamefont {T.}~\bibnamefont {Kachel}}, \bibinfo {author}
  {\bibfnamefont {N.}~\bibnamefont {Pontius}}, \bibinfo {author} {\bibfnamefont
  {H.~A.}\ \bibnamefont {D{\"u}rr}}, \bibinfo {author} {\bibfnamefont {T.~A.}\
  \bibnamefont {Ostler}}, \bibinfo {author} {\bibfnamefont {J.}~\bibnamefont
  {Barker}}, \bibinfo {author} {\bibfnamefont {R.~F.~L.}\ \bibnamefont
  {Evans}}, \bibinfo {author} {\bibfnamefont {R.~W.}\ \bibnamefont
  {Chantrell}}, \bibinfo {author} {\bibfnamefont {A.}~\bibnamefont
  {Tsukamoto}}, \bibinfo {author} {\bibfnamefont {A.}~\bibnamefont {Itoh}},
  \bibinfo {author} {\bibfnamefont {A.}~\bibnamefont {Kirilyuk}}, \bibinfo
  {author} {\bibfnamefont {T.}~\bibnamefont {Rasing}}, \ and\ \bibinfo {author}
  {\bibfnamefont {A.~V.}\ \bibnamefont {Kimel}},\ }\href {\doibase
  10.1038/nature09901} {\bibfield  {journal} {\bibinfo  {journal} {Nature}\
  }\textbf {\bibinfo {volume} {472}},\ \bibinfo {pages} {205} (\bibinfo {year}
  {2011})}\BibitemShut {NoStop}%
\bibitem [{\citenamefont {Boeglin}\ \emph {et~al.}(2010)\citenamefont
  {Boeglin}, \citenamefont {Beaurepaire}, \citenamefont {Halt{\'e}},
  \citenamefont {L{\'o}pez-Flores}, \citenamefont {Stamm}, \citenamefont
  {Pontius}, \citenamefont {D{\"u}rr},\ and\ \citenamefont
  {Bigot}}]{Boeglin2010}%
  \BibitemOpen
  \bibfield  {author} {\bibinfo {author} {\bibfnamefont {C.}~\bibnamefont
  {Boeglin}}, \bibinfo {author} {\bibfnamefont {E.}~\bibnamefont
  {Beaurepaire}}, \bibinfo {author} {\bibfnamefont {V.}~\bibnamefont
  {Halt{\'e}}}, \bibinfo {author} {\bibfnamefont {V.}~\bibnamefont
  {L{\'o}pez-Flores}}, \bibinfo {author} {\bibfnamefont {C.}~\bibnamefont
  {Stamm}}, \bibinfo {author} {\bibfnamefont {N.}~\bibnamefont {Pontius}},
  \bibinfo {author} {\bibfnamefont {H.~A.}\ \bibnamefont {D{\"u}rr}}, \ and\
  \bibinfo {author} {\bibfnamefont {J.-Y.}\ \bibnamefont {Bigot}},\ }\href
  {\doibase 10.1038/nature09070} {\bibfield  {journal} {\bibinfo  {journal}
  {Nature}\ }\textbf {\bibinfo {volume} {465}},\ \bibinfo {pages} {458}
  (\bibinfo {year} {2010})}\BibitemShut {NoStop}%
\bibitem [{\citenamefont {Jal}\ \emph {et~al.}(2019)\citenamefont {Jal},
  \citenamefont {Makita}, \citenamefont {R\"osner}, \citenamefont {David},
  \citenamefont {Nolting}, \citenamefont {Raabe}, \citenamefont {Savchenko},
  \citenamefont {Kleibert}, \citenamefont {Capotondi}, \citenamefont
  {Pedersoli}, \citenamefont {Raimondi}, \citenamefont {Manfredda},
  \citenamefont {Nikolov}, \citenamefont {Liu}, \citenamefont {Merhe},
  \citenamefont {Jaouen}, \citenamefont {Gorchon}, \citenamefont {Malinowski},
  \citenamefont {Hehn}, \citenamefont {Vodungbo},\ and\ \citenamefont
  {L\"uning}}]{Jal2019}%
  \BibitemOpen
  \bibfield  {author} {\bibinfo {author} {\bibfnamefont {E.}~\bibnamefont
  {Jal}}, \bibinfo {author} {\bibfnamefont {M.}~\bibnamefont {Makita}},
  \bibinfo {author} {\bibfnamefont {B.}~\bibnamefont {R\"osner}}, \bibinfo
  {author} {\bibfnamefont {C.}~\bibnamefont {David}}, \bibinfo {author}
  {\bibfnamefont {F.}~\bibnamefont {Nolting}}, \bibinfo {author} {\bibfnamefont
  {J.}~\bibnamefont {Raabe}}, \bibinfo {author} {\bibfnamefont
  {T.}~\bibnamefont {Savchenko}}, \bibinfo {author} {\bibfnamefont
  {A.}~\bibnamefont {Kleibert}}, \bibinfo {author} {\bibfnamefont
  {F.}~\bibnamefont {Capotondi}}, \bibinfo {author} {\bibfnamefont
  {E.}~\bibnamefont {Pedersoli}}, \bibinfo {author} {\bibfnamefont
  {L.}~\bibnamefont {Raimondi}}, \bibinfo {author} {\bibfnamefont
  {M.}~\bibnamefont {Manfredda}}, \bibinfo {author} {\bibfnamefont
  {I.}~\bibnamefont {Nikolov}}, \bibinfo {author} {\bibfnamefont
  {X.}~\bibnamefont {Liu}}, \bibinfo {author} {\bibfnamefont {A.~e.~d.}\
  \bibnamefont {Merhe}}, \bibinfo {author} {\bibfnamefont {N.}~\bibnamefont
  {Jaouen}}, \bibinfo {author} {\bibfnamefont {J.}~\bibnamefont {Gorchon}},
  \bibinfo {author} {\bibfnamefont {G.}~\bibnamefont {Malinowski}}, \bibinfo
  {author} {\bibfnamefont {M.}~\bibnamefont {Hehn}}, \bibinfo {author}
  {\bibfnamefont {B.}~\bibnamefont {Vodungbo}}, \ and\ \bibinfo {author}
  {\bibfnamefont {J.}~\bibnamefont {L\"uning}},\ }\href {\doibase
  10.1103/PhysRevB.99.144305} {\bibfield  {journal} {\bibinfo  {journal} {Phys.
  Rev. B}\ }\textbf {\bibinfo {volume} {99}},\ \bibinfo {pages} {144305}
  (\bibinfo {year} {2019})}\BibitemShut {NoStop}%
\bibitem [{\citenamefont {Radu}\ \emph {et~al.}(2015)\citenamefont {Radu},
  \citenamefont {Stamm}, \citenamefont {Eschenlohr}, \citenamefont {Radu},
  \citenamefont {Abrudan}, \citenamefont {Vahaplar}, \citenamefont {Kachel},
  \citenamefont {Pontius}, \citenamefont {Mitzner}, \citenamefont {Holldack},
  \citenamefont {Föhlisch}, \citenamefont {Ostler}, \citenamefont {Mentink},
  \citenamefont {Evans}, \citenamefont {Chantrell}, \citenamefont {Tsukamoto},
  \citenamefont {Itoh}, \citenamefont {Kirilyuk}, \citenamefont {Kimel},\ and\
  \citenamefont {Rasing}}]{Radu2015}%
  \BibitemOpen
  \bibfield  {author} {\bibinfo {author} {\bibfnamefont {I.}~\bibnamefont
  {Radu}}, \bibinfo {author} {\bibfnamefont {C.}~\bibnamefont {Stamm}},
  \bibinfo {author} {\bibfnamefont {A.}~\bibnamefont {Eschenlohr}}, \bibinfo
  {author} {\bibfnamefont {F.}~\bibnamefont {Radu}}, \bibinfo {author}
  {\bibfnamefont {R.}~\bibnamefont {Abrudan}}, \bibinfo {author} {\bibfnamefont
  {K.}~\bibnamefont {Vahaplar}}, \bibinfo {author} {\bibfnamefont
  {T.}~\bibnamefont {Kachel}}, \bibinfo {author} {\bibfnamefont
  {N.}~\bibnamefont {Pontius}}, \bibinfo {author} {\bibfnamefont
  {R.}~\bibnamefont {Mitzner}}, \bibinfo {author} {\bibfnamefont
  {K.}~\bibnamefont {Holldack}}, \bibinfo {author} {\bibfnamefont
  {A.}~\bibnamefont {Föhlisch}}, \bibinfo {author} {\bibfnamefont {T.~A.}\
  \bibnamefont {Ostler}}, \bibinfo {author} {\bibfnamefont {J.~H.}\
  \bibnamefont {Mentink}}, \bibinfo {author} {\bibfnamefont {R.~F.~L.}\
  \bibnamefont {Evans}}, \bibinfo {author} {\bibfnamefont {R.~W.}\ \bibnamefont
  {Chantrell}}, \bibinfo {author} {\bibfnamefont {A.}~\bibnamefont
  {Tsukamoto}}, \bibinfo {author} {\bibfnamefont {A.}~\bibnamefont {Itoh}},
  \bibinfo {author} {\bibfnamefont {A.}~\bibnamefont {Kirilyuk}}, \bibinfo
  {author} {\bibfnamefont {A.~V.}\ \bibnamefont {Kimel}}, \ and\ \bibinfo
  {author} {\bibfnamefont {T.}~\bibnamefont {Rasing}},\ }\href
  {https://doi.org/10.1142/S2010324715500046} {\bibfield  {journal} {\bibinfo
  {journal} {SPIN}\ }\textbf {\bibinfo {volume} {05}},\ \bibinfo {pages}
  {1550004} (\bibinfo {year} {2015})}\BibitemShut {NoStop}%
\bibitem [{\citenamefont {Sant}\ \emph {et~al.}(2017)\citenamefont {Sant},
  \citenamefont {Ksenzov}, \citenamefont {Capotondi}, \citenamefont
  {Pedersoli}, \citenamefont {Manfredda}, \citenamefont {Kiskinova},
  \citenamefont {Zabel}, \citenamefont {Kl{\"a}ui}, \citenamefont {L{\"u}ning},
  \citenamefont {Pietsch},\ and\ \citenamefont {Gutt}}]{Sant2017}%
  \BibitemOpen
  \bibfield  {author} {\bibinfo {author} {\bibfnamefont {T.}~\bibnamefont
  {Sant}}, \bibinfo {author} {\bibfnamefont {D.}~\bibnamefont {Ksenzov}},
  \bibinfo {author} {\bibfnamefont {F.}~\bibnamefont {Capotondi}}, \bibinfo
  {author} {\bibfnamefont {E.}~\bibnamefont {Pedersoli}}, \bibinfo {author}
  {\bibfnamefont {M.}~\bibnamefont {Manfredda}}, \bibinfo {author}
  {\bibfnamefont {M.}~\bibnamefont {Kiskinova}}, \bibinfo {author}
  {\bibfnamefont {H.}~\bibnamefont {Zabel}}, \bibinfo {author} {\bibfnamefont
  {M.}~\bibnamefont {Kl{\"a}ui}}, \bibinfo {author} {\bibfnamefont
  {J.}~\bibnamefont {L{\"u}ning}}, \bibinfo {author} {\bibfnamefont
  {U.}~\bibnamefont {Pietsch}}, \ and\ \bibinfo {author} {\bibfnamefont
  {C.}~\bibnamefont {Gutt}},\ }\href {\doibase 10.1038/s41598-017-15234-7}
  {\bibfield  {journal} {\bibinfo  {journal} {Scientific Reports}\ }\textbf
  {\bibinfo {volume} {7}},\ \bibinfo {pages} {15064} (\bibinfo {year}
  {2017})}\BibitemShut {NoStop}%
\bibitem [{\citenamefont {O'Handley}(2000)}]{OHandleyBOOK}%
  \BibitemOpen
  \bibfield  {author} {\bibinfo {author} {\bibfnamefont {R.~C.}\ \bibnamefont
  {O'Handley}},\ }\href@noop {} {\emph {\bibinfo {title} {Modern {M}agnetic
  {M}aterials}}}\ (\bibinfo  {publisher} {Wiley},\ \bibinfo {address} {New
  York},\ \bibinfo {year} {c2000})\BibitemShut {NoStop}%
\bibitem [{\citenamefont {Pudell}\ \emph {et~al.}(2018)\citenamefont {Pudell},
  \citenamefont {Maznev}, \citenamefont {Herzog}, \citenamefont {Kronseder},
  \citenamefont {Back}, \citenamefont {Malinowski}, \citenamefont {von
  Reppert},\ and\ \citenamefont {Bargheer}}]{Pudell2018}%
  \BibitemOpen
  \bibfield  {author} {\bibinfo {author} {\bibfnamefont {J.}~\bibnamefont
  {Pudell}}, \bibinfo {author} {\bibfnamefont {A.~A.}\ \bibnamefont {Maznev}},
  \bibinfo {author} {\bibfnamefont {M.}~\bibnamefont {Herzog}}, \bibinfo
  {author} {\bibfnamefont {M.}~\bibnamefont {Kronseder}}, \bibinfo {author}
  {\bibfnamefont {C.~H.}\ \bibnamefont {Back}}, \bibinfo {author}
  {\bibfnamefont {G.}~\bibnamefont {Malinowski}}, \bibinfo {author}
  {\bibfnamefont {A.}~\bibnamefont {von Reppert}}, \ and\ \bibinfo {author}
  {\bibfnamefont {M.}~\bibnamefont {Bargheer}},\ }\href {\doibase
  10.1038/s41467-018-05693-5} {\bibfield  {journal} {\bibinfo  {journal}
  {Nature Communications}\ }\textbf {\bibinfo {volume} {9}},\ \bibinfo {pages}
  {3335} (\bibinfo {year} {2018})}\BibitemShut {NoStop}%
\bibitem [{\citenamefont {Maldonado}\ \emph {et~al.}(2020)\citenamefont
  {Maldonado}, \citenamefont {Chase}, \citenamefont {Reid}, \citenamefont
  {Shen}, \citenamefont {Li}, \citenamefont {Carva}, \citenamefont {Payer},
  \citenamefont {Horn~von Hoegen}, \citenamefont {Sokolowski-Tinten},
  \citenamefont {Wang}, \citenamefont {Oppeneer},\ and\ \citenamefont
  {D\"urr}}]{Maldonado2020}%
  \BibitemOpen
  \bibfield  {author} {\bibinfo {author} {\bibfnamefont {P.}~\bibnamefont
  {Maldonado}}, \bibinfo {author} {\bibfnamefont {T.}~\bibnamefont {Chase}},
  \bibinfo {author} {\bibfnamefont {A.~H.}\ \bibnamefont {Reid}}, \bibinfo
  {author} {\bibfnamefont {X.}~\bibnamefont {Shen}}, \bibinfo {author}
  {\bibfnamefont {R.~K.}\ \bibnamefont {Li}}, \bibinfo {author} {\bibfnamefont
  {K.}~\bibnamefont {Carva}}, \bibinfo {author} {\bibfnamefont
  {T.}~\bibnamefont {Payer}}, \bibinfo {author} {\bibfnamefont
  {M.}~\bibnamefont {Horn~von Hoegen}}, \bibinfo {author} {\bibfnamefont
  {K.}~\bibnamefont {Sokolowski-Tinten}}, \bibinfo {author} {\bibfnamefont
  {X.~J.}\ \bibnamefont {Wang}}, \bibinfo {author} {\bibfnamefont {P.~M.}\
  \bibnamefont {Oppeneer}}, \ and\ \bibinfo {author} {\bibfnamefont {H.~A.}\
  \bibnamefont {D\"urr}},\ }\href {\doibase 10.1103/PhysRevB.101.100302}
  {\bibfield  {journal} {\bibinfo  {journal} {Phys. Rev. B}\ }\textbf {\bibinfo
  {volume} {101}},\ \bibinfo {pages} {100302} (\bibinfo {year}
  {2020})}\BibitemShut {NoStop}%
\bibitem [{\citenamefont {Hansen}\ \emph {et~al.}(1991)\citenamefont {Hansen},
  \citenamefont {Klahn}, \citenamefont {Clausen}, \citenamefont {Much},\ and\
  \citenamefont {Witter}}]{Witter1991}%
  \BibitemOpen
  \bibfield  {author} {\bibinfo {author} {\bibfnamefont {P.}~\bibnamefont
  {Hansen}}, \bibinfo {author} {\bibfnamefont {S.}~\bibnamefont {Klahn}},
  \bibinfo {author} {\bibfnamefont {C.}~\bibnamefont {Clausen}}, \bibinfo
  {author} {\bibfnamefont {G.}~\bibnamefont {Much}}, \ and\ \bibinfo {author}
  {\bibfnamefont {K.}~\bibnamefont {Witter}},\ }\href
  {https://doi.org/10.1063/1.348561} {\bibfield  {journal} {\bibinfo  {journal}
  {Journal of Applied Physics}\ }\textbf {\bibinfo {volume} {69}},\ \bibinfo
  {pages} {3194} (\bibinfo {year} {1991})}\BibitemShut {NoStop}%
\bibitem [{\citenamefont {Moreno}\ \emph {et~al.}(2016)\citenamefont {Moreno},
  \citenamefont {Evans}, \citenamefont {Khmelevskyi}, \citenamefont {Mu\~noz},
  \citenamefont {Chantrell},\ and\ \citenamefont
  {Chubykalo-Fesenko}}]{Fesenko2016}%
  \BibitemOpen
  \bibfield  {author} {\bibinfo {author} {\bibfnamefont {R.}~\bibnamefont
  {Moreno}}, \bibinfo {author} {\bibfnamefont {R.~F.~L.}\ \bibnamefont
  {Evans}}, \bibinfo {author} {\bibfnamefont {S.}~\bibnamefont {Khmelevskyi}},
  \bibinfo {author} {\bibfnamefont {M.~C.}\ \bibnamefont {Mu\~noz}}, \bibinfo
  {author} {\bibfnamefont {R.~W.}\ \bibnamefont {Chantrell}}, \ and\ \bibinfo
  {author} {\bibfnamefont {O.}~\bibnamefont {Chubykalo-Fesenko}},\ }\href
  {\doibase 10.1103/PhysRevB.94.104433} {\bibfield  {journal} {\bibinfo
  {journal} {Phys. Rev. B}\ }\textbf {\bibinfo {volume} {94}},\ \bibinfo
  {pages} {104433} (\bibinfo {year} {2016})}\BibitemShut {NoStop}%
\bibitem [{\citenamefont {Virot}\ \emph {et~al.}(2012)\citenamefont {Virot},
  \citenamefont {Favre}, \citenamefont {Hayn},\ and\ \citenamefont
  {Kuz{\textquotesingle}min}}]{Virot2012}%
  \BibitemOpen
  \bibfield  {author} {\bibinfo {author} {\bibfnamefont {F.}~\bibnamefont
  {Virot}}, \bibinfo {author} {\bibfnamefont {L.}~\bibnamefont {Favre}},
  \bibinfo {author} {\bibfnamefont {R.}~\bibnamefont {Hayn}}, \ and\ \bibinfo
  {author} {\bibfnamefont {M.~D.}\ \bibnamefont {Kuz{\textquotesingle}min}},\
  }\href {\doibase 10.1088/0022-3727/45/40/405003} {\bibfield  {journal}
  {\bibinfo  {journal} {Journal of Physics D: Applied Physics}\ }\textbf
  {\bibinfo {volume} {45}},\ \bibinfo {pages} {405003} (\bibinfo {year}
  {2012})}\BibitemShut {NoStop}%
\bibitem [{\citenamefont {El~Hadri}\ \emph
  {et~al.}(2016{\natexlab{b}})\citenamefont {El~Hadri}, \citenamefont {Hehn},
  \citenamefont {Pirro}, \citenamefont {Lambert}, \citenamefont {Malinowski},
  \citenamefont {Fullerton},\ and\ \citenamefont {Mangin}}]{ElHadri2016}%
  \BibitemOpen
  \bibfield  {author} {\bibinfo {author} {\bibfnamefont {M.~S.}\ \bibnamefont
  {El~Hadri}}, \bibinfo {author} {\bibfnamefont {M.}~\bibnamefont {Hehn}},
  \bibinfo {author} {\bibfnamefont {P.}~\bibnamefont {Pirro}}, \bibinfo
  {author} {\bibfnamefont {C.-H.}\ \bibnamefont {Lambert}}, \bibinfo {author}
  {\bibfnamefont {G.}~\bibnamefont {Malinowski}}, \bibinfo {author}
  {\bibfnamefont {E.~E.}\ \bibnamefont {Fullerton}}, \ and\ \bibinfo {author}
  {\bibfnamefont {S.}~\bibnamefont {Mangin}},\ }\href {\doibase
  10.1103/PhysRevB.94.064419} {\bibfield  {journal} {\bibinfo  {journal} {Phys.
  Rev. B}\ }\textbf {\bibinfo {volume} {94}},\ \bibinfo {pages} {064419}
  (\bibinfo {year} {2016}{\natexlab{b}})}\BibitemShut {NoStop}%
\bibitem [{\citenamefont {Kooy}\ and\ \citenamefont {Enz}(1960)}]{Kooy1960}%
  \BibitemOpen
  \bibfield  {author} {\bibinfo {author} {\bibfnamefont {C.}~\bibnamefont
  {Kooy}}\ and\ \bibinfo {author} {\bibfnamefont {U.}~\bibnamefont {Enz}},\
  }\href@noop {} {\bibfield  {journal} {\bibinfo  {journal} {Philips Res.
  Rep.}\ }\textbf {\bibinfo {volume} {15}} (\bibinfo {year}
  {1960})}\BibitemShut {NoStop}%
\bibitem [{\citenamefont {Turner}\ \emph {et~al.}(2011)\citenamefont {Turner},
  \citenamefont {Huang}, \citenamefont {Krupin}, \citenamefont {Seu},
  \citenamefont {Parks}, \citenamefont {Kevan}, \citenamefont {Lima},
  \citenamefont {Kisslinger}, \citenamefont {McNulty}, \citenamefont {Gambino},
  \citenamefont {Mangin}, \citenamefont {Roy},\ and\ \citenamefont
  {Fischer}}]{Turner2011}%
  \BibitemOpen
  \bibfield  {author} {\bibinfo {author} {\bibfnamefont {J.~J.}\ \bibnamefont
  {Turner}}, \bibinfo {author} {\bibfnamefont {X.}~\bibnamefont {Huang}},
  \bibinfo {author} {\bibfnamefont {O.}~\bibnamefont {Krupin}}, \bibinfo
  {author} {\bibfnamefont {K.~A.}\ \bibnamefont {Seu}}, \bibinfo {author}
  {\bibfnamefont {D.}~\bibnamefont {Parks}}, \bibinfo {author} {\bibfnamefont
  {S.}~\bibnamefont {Kevan}}, \bibinfo {author} {\bibfnamefont
  {E.}~\bibnamefont {Lima}}, \bibinfo {author} {\bibfnamefont {K.}~\bibnamefont
  {Kisslinger}}, \bibinfo {author} {\bibfnamefont {I.}~\bibnamefont {McNulty}},
  \bibinfo {author} {\bibfnamefont {R.}~\bibnamefont {Gambino}}, \bibinfo
  {author} {\bibfnamefont {S.}~\bibnamefont {Mangin}}, \bibinfo {author}
  {\bibfnamefont {S.}~\bibnamefont {Roy}}, \ and\ \bibinfo {author}
  {\bibfnamefont {P.}~\bibnamefont {Fischer}},\ }\href {\doibase
  10.1103/PhysRevLett.107.033904} {\bibfield  {journal} {\bibinfo  {journal}
  {Phys. Rev. Lett.}\ }\textbf {\bibinfo {volume} {107}},\ \bibinfo {pages}
  {033904} (\bibinfo {year} {2011})}\BibitemShut {NoStop}%
\bibitem [{\citenamefont {Samsonov}(1968)}]{Samsonov1968}%
  \BibitemOpen
  \bibinfo {editor} {\bibfnamefont {G.~V.}\ \bibnamefont {Samsonov}},\ ed.,\
  \href@noop {} {\emph {\bibinfo {title} {Handbook of the Physicochemical
  Properties of the Elements}}}\ (\bibinfo  {publisher} {New York: Plenum
  Publishing Corporation},\ \bibinfo {year} {1968})\BibitemShut {NoStop}%
\bibitem [{\citenamefont {Bang}\ \emph {et~al.}(2016)\citenamefont {Bang},
  \citenamefont {Yu}, \citenamefont {Qiu}, \citenamefont {Wang}, \citenamefont
  {Awano}, \citenamefont {Manchon},\ and\ \citenamefont {Yang}}]{Bang2016}%
  \BibitemOpen
  \bibfield  {author} {\bibinfo {author} {\bibfnamefont {D.}~\bibnamefont
  {Bang}}, \bibinfo {author} {\bibfnamefont {J.}~\bibnamefont {Yu}}, \bibinfo
  {author} {\bibfnamefont {X.}~\bibnamefont {Qiu}}, \bibinfo {author}
  {\bibfnamefont {Y.}~\bibnamefont {Wang}}, \bibinfo {author} {\bibfnamefont
  {H.}~\bibnamefont {Awano}}, \bibinfo {author} {\bibfnamefont
  {A.}~\bibnamefont {Manchon}}, \ and\ \bibinfo {author} {\bibfnamefont
  {H.}~\bibnamefont {Yang}},\ }\href {\doibase 10.1103/PhysRevB.93.174424}
  {\bibfield  {journal} {\bibinfo  {journal} {Phys. Rev. B}\ }\textbf {\bibinfo
  {volume} {93}},\ \bibinfo {pages} {174424} (\bibinfo {year}
  {2016})}\BibitemShut {NoStop}%
\bibitem [{\citenamefont {Siddiqui}\ \emph {et~al.}(2018)\citenamefont
  {Siddiqui}, \citenamefont {Han}, \citenamefont {Finley}, \citenamefont
  {Ross},\ and\ \citenamefont {Liu}}]{Ross2018}%
  \BibitemOpen
  \bibfield  {author} {\bibinfo {author} {\bibfnamefont {S.~A.}\ \bibnamefont
  {Siddiqui}}, \bibinfo {author} {\bibfnamefont {J.}~\bibnamefont {Han}},
  \bibinfo {author} {\bibfnamefont {J.~T.}\ \bibnamefont {Finley}}, \bibinfo
  {author} {\bibfnamefont {C.~A.}\ \bibnamefont {Ross}}, \ and\ \bibinfo
  {author} {\bibfnamefont {L.}~\bibnamefont {Liu}},\ }\href {\doibase
  10.1103/PhysRevLett.121.057701} {\bibfield  {journal} {\bibinfo  {journal}
  {Phys. Rev. Lett.}\ }\textbf {\bibinfo {volume} {121}},\ \bibinfo {pages}
  {057701} (\bibinfo {year} {2018})}\BibitemShut {NoStop}%
\bibitem [{\citenamefont {Dalla~Longa}\ \emph {et~al.}(2007)\citenamefont
  {Dalla~Longa}, \citenamefont {Kohlhepp}, \citenamefont {de~Jonge},\ and\
  \citenamefont {Koopmans}}]{DallaLonga2007}%
  \BibitemOpen
  \bibfield  {author} {\bibinfo {author} {\bibfnamefont {F.}~\bibnamefont
  {Dalla~Longa}}, \bibinfo {author} {\bibfnamefont {J.~T.}\ \bibnamefont
  {Kohlhepp}}, \bibinfo {author} {\bibfnamefont {W.~J.~M.}\ \bibnamefont
  {de~Jonge}}, \ and\ \bibinfo {author} {\bibfnamefont {B.}~\bibnamefont
  {Koopmans}},\ }\href {\doibase 10.1103/PhysRevB.75.224431} {\bibfield
  {journal} {\bibinfo  {journal} {Phys. Rev. B}\ }\textbf {\bibinfo {volume}
  {75}},\ \bibinfo {pages} {224431} (\bibinfo {year} {2007})}\BibitemShut
  {NoStop}%
\bibitem [{\citenamefont {Moisan}\ \emph
  {et~al.}(2014{\natexlab{b}})\citenamefont {Moisan}, \citenamefont
  {Malinowski}, \citenamefont {Mauchain}, \citenamefont {Hehn}, \citenamefont
  {Vodungbo}, \citenamefont {L{\"u}ning}, \citenamefont {Mangin}, \citenamefont
  {Fullerton},\ and\ \citenamefont {Thiaville}}]{Moisan2015}%
  \BibitemOpen
  \bibfield  {author} {\bibinfo {author} {\bibfnamefont {N.}~\bibnamefont
  {Moisan}}, \bibinfo {author} {\bibfnamefont {G.}~\bibnamefont {Malinowski}},
  \bibinfo {author} {\bibfnamefont {J.}~\bibnamefont {Mauchain}}, \bibinfo
  {author} {\bibfnamefont {M.}~\bibnamefont {Hehn}}, \bibinfo {author}
  {\bibfnamefont {B.}~\bibnamefont {Vodungbo}}, \bibinfo {author}
  {\bibfnamefont {J.}~\bibnamefont {L{\"u}ning}}, \bibinfo {author}
  {\bibfnamefont {S.}~\bibnamefont {Mangin}}, \bibinfo {author} {\bibfnamefont
  {E.~E.}\ \bibnamefont {Fullerton}}, \ and\ \bibinfo {author} {\bibfnamefont
  {A.}~\bibnamefont {Thiaville}},\ }\href {https://doi.org/10.1038/srep04658}
  {\bibfield  {journal} {\bibinfo  {journal} {Scientific Reports}\ }\textbf
  {\bibinfo {volume} {4}},\ \bibinfo {pages} {4658 EP } (\bibinfo {year}
  {2014}{\natexlab{b}})}\BibitemShut {NoStop}%
\bibitem [{\citenamefont {Vodungbo}\ \emph {et~al.}(2016)\citenamefont
  {Vodungbo}, \citenamefont {Tudu}, \citenamefont {Perron}, \citenamefont
  {Delaunay}, \citenamefont {M{\"u}ller}, \citenamefont {Berntsen},
  \citenamefont {Gr{\"u}bel}, \citenamefont {Malinowski}, \citenamefont
  {Weier}, \citenamefont {Gautier}, \citenamefont {Lambert}, \citenamefont
  {Zeitoun}, \citenamefont {Gutt}, \citenamefont {Jal}, \citenamefont {Reid},
  \citenamefont {Granitzka}, \citenamefont {Jaouen}, \citenamefont {Dakovski},
  \citenamefont {Moeller}, \citenamefont {Minitti}, \citenamefont {Mitra},
  \citenamefont {Carron}, \citenamefont {Pfau}, \citenamefont {von
  Korff~Schmising}, \citenamefont {Schneider}, \citenamefont {Eisebitt},\ and\
  \citenamefont {L{\"u}ning}}]{Vodungbo2016}%
  \BibitemOpen
  \bibfield  {author} {\bibinfo {author} {\bibfnamefont {B.}~\bibnamefont
  {Vodungbo}}, \bibinfo {author} {\bibfnamefont {B.}~\bibnamefont {Tudu}},
  \bibinfo {author} {\bibfnamefont {J.}~\bibnamefont {Perron}}, \bibinfo
  {author} {\bibfnamefont {R.}~\bibnamefont {Delaunay}}, \bibinfo {author}
  {\bibfnamefont {L.}~\bibnamefont {M{\"u}ller}}, \bibinfo {author}
  {\bibfnamefont {M.~H.}\ \bibnamefont {Berntsen}}, \bibinfo {author}
  {\bibfnamefont {G.}~\bibnamefont {Gr{\"u}bel}}, \bibinfo {author}
  {\bibfnamefont {G.}~\bibnamefont {Malinowski}}, \bibinfo {author}
  {\bibfnamefont {C.}~\bibnamefont {Weier}}, \bibinfo {author} {\bibfnamefont
  {J.}~\bibnamefont {Gautier}}, \bibinfo {author} {\bibfnamefont
  {G.}~\bibnamefont {Lambert}}, \bibinfo {author} {\bibfnamefont
  {P.}~\bibnamefont {Zeitoun}}, \bibinfo {author} {\bibfnamefont
  {C.}~\bibnamefont {Gutt}}, \bibinfo {author} {\bibfnamefont {E.}~\bibnamefont
  {Jal}}, \bibinfo {author} {\bibfnamefont {A.~H.}\ \bibnamefont {Reid}},
  \bibinfo {author} {\bibfnamefont {P.~W.}\ \bibnamefont {Granitzka}}, \bibinfo
  {author} {\bibfnamefont {N.}~\bibnamefont {Jaouen}}, \bibinfo {author}
  {\bibfnamefont {G.~L.}\ \bibnamefont {Dakovski}}, \bibinfo {author}
  {\bibfnamefont {S.}~\bibnamefont {Moeller}}, \bibinfo {author} {\bibfnamefont
  {M.~P.}\ \bibnamefont {Minitti}}, \bibinfo {author} {\bibfnamefont
  {A.}~\bibnamefont {Mitra}}, \bibinfo {author} {\bibfnamefont
  {S.}~\bibnamefont {Carron}}, \bibinfo {author} {\bibfnamefont
  {B.}~\bibnamefont {Pfau}}, \bibinfo {author} {\bibfnamefont {C.}~\bibnamefont
  {von Korff~Schmising}}, \bibinfo {author} {\bibfnamefont {M.}~\bibnamefont
  {Schneider}}, \bibinfo {author} {\bibfnamefont {S.}~\bibnamefont {Eisebitt}},
  \ and\ \bibinfo {author} {\bibfnamefont {J.}~\bibnamefont {L{\"u}ning}},\
  }\href {https://doi.org/10.1038/srep18970} {\bibfield  {journal} {\bibinfo
  {journal} {Scientific Reports}\ }\textbf {\bibinfo {volume} {6}},\ \bibinfo
  {pages} {18970 EP } (\bibinfo {year} {2016})}\BibitemShut {NoStop}%
\bibitem [{\citenamefont {Hasegawa}(1974)}]{Hasegawa1974}%
  \BibitemOpen
  \bibfield  {author} {\bibinfo {author} {\bibfnamefont {R.}~\bibnamefont
  {Hasegawa}},\ }\href {https://doi.org/10.1063/1.1663732} {\bibfield
  {journal} {\bibinfo  {journal} {Journal of Applied Physics}\ }\textbf
  {\bibinfo {volume} {45}},\ \bibinfo {pages} {3109} (\bibinfo {year}
  {1974})}\BibitemShut {NoStop}%
\bibitem [{\citenamefont {Zou}\ \emph {et~al.}(2003)\citenamefont {Zou},
  \citenamefont {Wang},\ and\ \citenamefont {Yu}}]{Zou2003}%
  \BibitemOpen
  \bibfield  {author} {\bibinfo {author} {\bibfnamefont {Z.~Q.}\ \bibnamefont
  {Zou}}, \bibinfo {author} {\bibfnamefont {H.}~\bibnamefont {Wang}}, \ and\
  \bibinfo {author} {\bibfnamefont {C.}~\bibnamefont {Yu}},\ }\href
  {https://doi.org/10.1063/1.1565828} {\bibfield  {journal} {\bibinfo
  {journal} {Journal of Applied Physics}\ }\textbf {\bibinfo {volume} {93}},\
  \bibinfo {pages} {5268} (\bibinfo {year} {2003})}\BibitemShut {NoStop}%
\end{thebibliography}%

\end{document}